\documentclass[nofootinbib,twocolumn,preprintnumbers]{revtex4-1}
\pdfoutput=1
\usepackage{amsmath,amsthm,amssymb,multirow,psfrag}
\usepackage{epsfig}
\usepackage{color}

\graphicspath{{./Figures/}}

\begin{document}

\def\lsim{\mathrel{\rlap{\lower4pt\hbox{\hskip1pt$\sim$}}
    \raise1pt\hbox{$<$}}}
\def\gsim{\mathrel{\rlap{\lower4pt\hbox{\hskip1pt$\sim$}}
    \raise1pt\hbox{$>$}}}
\newcommand{\vev}[1]{ \left\langle {#1} \right\rangle }
\newcommand{\bra}[1]{ \langle {#1} | }
\newcommand{\ket}[1]{ | {#1} \rangle }
\newcommand{\ev}{ {\rm eV} }
\newcommand{\kev}{{\rm keV}}
\newcommand{\mev}{{\rm MeV}}
\newcommand{\gev}{{\rm GeV}}
\newcommand{\tev}{{\rm TeV}}
\newcommand{\mpl}{$M_{Pl}$}
\newcommand{\mw}{$M_{W}$}
\newcommand{\Ft}{F_{T}}
\newcommand{\Zparity}{\mathbb{Z}_2}
\newcommand{\BLambda}{\boldsymbol{\lambda}}
\newcommand{\bea}{\begin{eqnarray}}
\newcommand{\eea}{\end{eqnarray}}
\newcommand{\met}{\;\not\!\!\!{E}_T}
\newcommand{\draftnote}[1]{{\bf\color{blue} #1}}

\newcommand{\fref}[1]{Fig.~\ref{fig:#1}} 
\newcommand{\eref}[1]{Eq.~\eqref{eqn:#1}} 
\newcommand{\aref}[1]{Appendix~\ref{app:#1}}
\newcommand{\sref}[1]{Sec.~\ref{sec:#1}}
\newcommand{\ssref}[1]{Sec.~\ref{subsec:#1}}
\newcommand{\tref}[1]{Table~\ref{tab:#1}}

\title{{\LARGE{\bf New Opportunities in $h\to 4\ell$}}}
\author{\bf{Yi Chen$\,^{a,\dagger}$,~Roni Harnik$\,^{b,\ddag}$,\,~Roberto Vega-Morales$\,^{c,\#}$}}

\affiliation{
$^a$Lauritsen Laboratory for High Energy Physics, California Institute of Technology, Pasadena, CA, 92115,\\
$^b$Theoretical Physics Department, Fermilab, P.O. Box 500, Batavia, IL 60510, USA,\\
$^c$Laboratoire de Physique Th\'{e}orique d'Orsay, UMR8627-CNRS, Universit\'{e} Paris-Sud, Orsay, France}

\email{$^\dagger$yichen@caltech.edu\\
$^\ddag$roni@fnal.gov\\
$^\#$roberto.vega@th.u-psud.fr}

\begin{abstract}
The Higgs decay $h\to 4\ell$ has played an important role in discovering the Higgs and measuring its mass thanks to low background and excellent resolution.~Current cuts in this channel have been optimized for Higgs discovery via the dominant tree level $ZZ$ contribution arising from electroweak symmetry breaking.~Going forward, one of the primary objectives of this sensitive channel will be to probe other Higgs couplings and search for new physics on top of the tree level $ZZ$ `background'.~Thanks to interference between these small couplings and the large tree level contribution to $ZZ$, the $h\to 4\ell$ decay is uniquely capable of probing the magnitude \emph{and} CP phases of the Higgs couplings to $\gamma\gamma$ and $Z\gamma$ as well as, to a lesser extent, $ZZ$ couplings arising from higher dimensional operators.~With this in mind we examine how much relaxing current cuts can enhance the sensitivity while also accounting for the dominant non-Higgs continuum $q\bar{q}\to4\ell$ background.~We find the largest enhancement in sensitivity for the $hZ\gamma$ couplings ($\gtrsim 100\%$) followed by $h\gamma\gamma$ ($\gtrsim 40\%$) and less so for the higher dimensional $hZZ$ couplings (a few percent).~With these enhancements, we show that couplings of order Standard Model values for $h\gamma\gamma$ may optimistically be probed by end of Run-II at the LHC while for $hZ\gamma$ perhaps towards the end of a high luminosity LHC.~Thus an appropriately optimized $h\to4\ell$ analysis can complement direct decays of the Higgs to on-shell $\gamma\gamma$ and $Z\gamma$ pairs giving a unique opportunity to directly access the CP properties of these couplings.

\end{abstract}

\preprint{FERMILAB-PUB-15-091-T, LPT-Orsay-15-22}

\maketitle

\section{Introduction}\label{sec:Intro}

The discovery of the Higgs boson at the LHC~\cite{:2012gk,:2012gu} has established that its properties closely resemble those predicted by the Standard Model (SM)~\cite{Falkowski:2013dza}.~The focus now shifts to the determination of its detailed properties and in particular whether or not it possesses any anomalous couplings not predicted by the SM.~It is thus important to re-examine current Higgs analyses with this shift in focus in mind.~In particular, analyses and cuts designed to discover the Higgs should now be optimized for more precise tests of Higgs couplings and searches for new physics. 

The decay of the Higgs to four leptons (electrons and muons) was one of the key channels in the discovery of the Higgs and the measurement of its mass.~This decay, which has been dubbed the `golden channel', has a small branching fraction, $\sim 10^{-4}$~in the SM, but this is compensated for by a high signal to background ratio as well as the high precision with which it is measured.~A small number of events, of order $\sim 10-15$ per experiment, were thus sufficient to claim discovery in the $h\to4\ell$ channel at both CMS and ATLAS~\cite{:2012gk,:2012gu}.

The $h\to4\ell$ decay (where $4\ell \equiv 2e2\mu, 4e, 4\mu$) is dominated by the $h\to ZZ$ component because of the large tree level coupling of the Higgs to $Z$ pairs which is generated by electroweak symmetry breaking (EWSB) in the SM and directly related to the way in which the $Z$ boson obtains its mass.~The cuts in the $h\to 4\ell$ analysis were thus designed to enhance this part of the amplitude over the continuum (mostly $q\bar{q}\to 4\ell$~\cite{CMS-PAS-HIG-14-014,Khachatryan:2014kca}) SM background.~However, with the establishment of a SM-like Higgs boson, the part of the $h\to4\ell$ decay which comes from the $hZ^\mu Z_\mu$ coupling should now be considered part of the SM \emph{background} and in fact, it composes the dominant background to the signal we are now after -- deviations from the standard model prediction for Higgs couplings.~One place such deviations can appear are the higher dimensional Higgs couplings to $ZZ, Z\gamma$, and $\gamma\gamma$ (we do not distinguish between on or off-shell) which contribute to the $h\to4\ell$ differential decay width.

Numerous studies have examined the golden channel as a probe of the Higgs couplings to $ZZ$ pairs including the CP properties at the LHC~\cite{Nelson:1986ki,Soni:1993jc,Chang:1993jy,Barger:1993wt,Arens:1994wd,Choi:2002jk,Buszello:2002uu,Godbole:2007cn,Kovalchuk:2008zz,Cao:2009ah,Gao:2010qx,DeRujula:2010ys,Gainer:2011xz,Coleppa:2012eh,Bolognesi:2012mm,Stolarski:2012ps,Boughezal:2012tz,Belyaev:2012qa,Avery:2012um,Campbell:2012ct,Campbell:2012cz,Chen:2012jy,Grinstein:2013vsa,Modak:2013sb,Sun:2013yra,Gainer:2013rxa,Anderson:2013fba,Chen:2013waa,Buchalla:2013mpa,Chen:2013ejz,Chen:2014pia,Gainer:2014hha}.~As we show below, since current cuts are optimized to uncover the tree level induced $h\to ZZ$ component, the sensitivity to the higher dimensional $ZZ$ operators is also already optimized.~We instead emphasize in this work the sensitivity of the golden channel to the higher-dimensional $h\gamma\gamma$ and $hZ\gamma$ couplings, which until recently~\cite{Chen:2012jy,Chen:2013ejz,Chen:2014pia,Chen:2014gka,Gonzalez-Alonso:2014eva} have been largely neglected in $h\to4\ell$ studies and only very recently studied experimentally for the first time by CMS~\cite{Khachatryan:2014kca}.~Our goal in this study is to assess the sensitivity to these couplings once the analysis is optimized for this purpose.

\section{Probing $hZ\gamma$ and $h\gamma\gamma$ Couplings\\~~~~in the Golden Channel}

One may wonder whether there is any advantage to searching for these couplings in $h\to4\ell$ rather than looking directly for Higgs decays to on-shell $\gamma\gamma$ and $Z\gamma$.~After all the rate to four lepton is suppressed by additional electroweak couplings and three or four-body phase space, compared to the two body phase space of direct decay to on-shell vector bosons.~Indeed, the coupling of the Higgs to photons is already well constrained by $h\to\gamma\gamma$.~There are a few important points to note when considering this:
\begin{itemize}
\item The signal rate in $h\to4\ell$ is indeed lower, but the backgrounds suffer from similar suppressions so the signal to background ratio is much larger~\cite{htoAA}.
\item The systematic uncertainties in the four lepton channel are very different than those in channels involving on-shell photons and typically smaller.
\item The large number of observables, of which there are twelve for the four massless fermions (see~\cite{Chen:2012jy,Chen:2013ejz,Chen:2014pia,Chen:2014gka} for a more detailed description), allows for better differentiation of signal versus background, almost on an event by event basis, especially in the case of the $\gamma\gamma$ contribution~\cite{Chen:2014gka}. 
\item Interference effects between the small $h\gamma\gamma$ and $hZ\gamma$ couplings with the large tree level $ZZ$ coupling allows the differential distributions to be sensitive to the CP phase of the respective couplings and possible CP violation~\cite{Chen:2014gka}.~Measurement of the rate into on-shell photons and $Z$'s is insensitive to CP violation\footnote{Note however it has been shown that sensitivity to CP violation is possible in the three-body $h\to2\ell\gamma$ decay~\cite{Chen:2014ona} when allowing for off-shell $Z$ and photon decays.~Also note that probing CPV by resolving converted photons is very challenging~\cite{Bishara:2013vya}.}.
\item The interference terms in the $h\to4\ell$ rate are proportional to the small higher dimensional couplings times the large $hZ_\mu Z^\mu$ coupling.~The rate into on-shell $\gamma$ and $Z$ goes like the small coupling squared.~Of course, interference terms are suppressed by other factors but this gives them a head-start in terms of sensitivity~\cite{Chen:2014gka}.
\end{itemize}
Indeed, it has been shown recently~\cite{Chen:2012jy,Chen:2013ejz,Chen:2014pia,Chen:2014gka} that the $h\to 4\ell$ ($4\ell \equiv 2e2\mu, 4e, 4\mu$) decay can be used to probe the Higgs couplings to $Z\gamma$ and $\gamma\gamma$ as well as $ZZ$ pairs.~In particular it was shown~\cite{Chen:2014gka} that even with existing cuts the LHC experiments will be able to probe sub-SM-sized~$h\gamma\gamma$ couplings by the end of high luminosity running while the sensitivity to SM-sized $hZ\gamma$ couplings is weaker, but possibly not hopeless.~This is despite the fact that these cuts were designed to enhance the Higgs discovery via the tree level $hZZ$ component.~In this work we examine relaxing some of the cuts in order to enhance the sensitivity to $h\gamma\gamma$ and $hZ\gamma$ couplings and assess to what extent the LHC may be able to probe these couplings.

\subsection{The $hVV$ Effective Couplings}

As in~\cite{Chen:2014gka} we consider the leading contributions to the Higgs couplings to neutral electroweak gauge bosons allowing for general CP odd/even mixtures as well as for $ZZ$, $Z\gamma$, and $\gamma\gamma$ to contribute simultaneously.~They can be parametrized by the following effective Lagrangian,
\begin{equation}
\label{eqn:fullL}
\mathcal{L}=\mathcal{L}_{o} + \mathcal{L}_{ZZ} 
+ \mathcal{L}_{Z\gamma} + \mathcal{L}_{\gamma\gamma} ,
\end{equation} 
where we have separated out the tree level term,
\bea
\label{eqn:A1}
\mathcal{L}_{o} &=& \frac{h}{2v} A_1^{ZZ} m_Z^2 Z^\mu Z_\mu .
\eea
This term is generated during EWSB and is responsible for giving the $Z$ boson its mass.~As in~\cite{Chen:2014gka} it will be treated as part of the background.~The higher dimensional `anomalous' operators in~\eref{fullL} are given by,
\bea
\label{eqn:dim5lag}
\mathcal{L}_{ZZ} &=& \frac{h}{4v}
\left(A_2^{ZZ} Z^{\mu\nu}Z_{\mu\nu} +
A_3^{ZZ} Z^{\mu\nu} \widetilde{Z}_{\mu\nu} \right) \nonumber  \\
\mathcal{L}_{Z\gamma} &=& \frac{h}{2v} 
\left(A_{2}^{Z\gamma} F^{\mu\nu}Z_{\mu\nu} 
+ A_3^{Z\gamma} F^{\mu\nu} \widetilde{Z}_{\mu\nu} \right) \\ 
\mathcal{L}_{\gamma\gamma} &=& \frac{h}{4v} 
\left(A_2^{\gamma\gamma} F^{\mu\nu}F_{\mu\nu} 
+ A_3^{\gamma\gamma} F^{\mu\nu} \widetilde{F}_{\mu\nu} \right) \nonumber ,
\eea
where all couplings are taken to be real, dimensionless, and constant.~Electromagnetic gauge invariance prohibits an $A_1$ type structure for the $Z\gamma$ and $\gamma\gamma$ couplings.

Note that strictly speaking our parametrization is not a completely general effective field theory (EFT) approach.~In a more general EFT approach one should also include other possible dimension five operators, such as $hZ_\mu\partial_\nu V^{\mu\nu}$ (where $V = Z,\gamma$) and $hZ_\mu \bar{\ell}\gamma^\mu \ell$ or $\Box h\, Z_\mu Z^\mu$ for off-shell Higgs decays~\cite{Gainer:2014hha} which we will not consider.~The interactions in~\eref{dim5lag} are thus just a representative set.~Furthermore, in the context of an underlying dimension six $SU(3)_c\times SU(2)_L\times U(1)_Y$ invariant EFT, correlations between these various operators are predicted~\cite{Pomarol:2013zra,Falkowski:2014tna}, but in the present study we treat them as independent.~Inclusion of these additional operators and studying their correlations in this context would be interesting as a means of probing regions of parameter space which are not constrained by LEP~\cite{Pomarol:2013zra,Falkowski:2014tna}, but we leave an exploration of this question to ongoing work~\cite{followup}.~We also note that in general the coefficients in~\eref{dim5lag} are momentum dependent form factors.~This is particularly true in the SM where the $W$ boson cannot be truly integrated out.~Still, to get an idea for the sensitivity of the $h\to4\ell$ search, it is sufficient to keep the leading order term in a momentum expansion, which is found to be dominant~\cite{Gonzalez-Alonso:2014eva,followup2}.

To summarize this discussion, our current focus is not to set precision constraints on possible higher dimensional operators~\cite{Elias-Miro:2013mua,Pomarol:2013zra,Croon:2014dma,Falkowski:2014tna} or define the optimal set of observables~\cite{Gupta:2014rxa,Trott:2014dma,Gonzalez-Alonso:2014eva}, but simply to establish at what point the LHC will begin to be sensitive to $\sim$ SM values of the $hZ\gamma$ and $h\gamma\gamma$ couplings defined in~\eref{dim5lag} if more optimized cuts are utilized.

\section{Constraints and Opportunities}

With these considerations in mind we follow the strategy presented in~\cite{Chen:2014gka} treating the $A_1^{ZZ}$ coupling as `background' and simultaneously fit for the other `loop induced' couplings in~\eref{dim5lag} to assess the sensitivity.~Thus, our six dimensional parameter space is defined as,
\bea
\label{eqn:Aall}
\vec{A} = (A^{ZZ}_2,A^{ZZ}_3,A^{Z\gamma}_2,A^{Z\gamma}_3,A^{\gamma\gamma}_2,A^{\gamma\gamma}_3) .
\eea
Detailed descriptions of the framework used for the parameter extraction and definitions of test statistics can be found in~\cite{Chen:2012jy,Chen:2013ejz,Chen:2014pia,Chen:2014gka}.~The couplings in~\eref{Aall} are currently constrained by LHC measurements and other experiments as follows:

\emph{\underline{Couplings to photons}}:~With the parametrization in~\eref{dim5lag} the SM values for the $\gamma\gamma$ couplings are $A^{\gamma\gamma}_2\sim-0.008$ and $A^{\gamma\gamma}_3\sim0$~\cite{Low:2012rj}.~The measurement of the Higgs signal strength in the diphoton channel by ATLAS and CMS places a constraint on the combination $|A^{\gamma\gamma}_2|^2 + |A^{\gamma\gamma}_3|^2$.~This combination is currently constrained to be about $1.55\pm0.3$ (ATLAS) and $0.77\pm0.27$ (CMS) times the SM value~\cite{htoAA,Aad:2013wqa}.~Note also that CMS has begun incorporating these couplings into their standard $h\to4\ell$ analysis of 7 and 8 TeV data~\cite{CMS-PAS-HIG-14-014, Khachatryan:2014kca}, but the sensitivity is still weak.~In addition, current limits on the electron electric dipole moment (EDM) require the CP phase $(A^{\gamma\gamma}_3/A^{\gamma\gamma}_2)$ to be very small, of order~$10^{-3}$~\cite{McKeen:2012av,Baron:2013eja}.~However, this limit is model dependent.~For example, if the 125 GeV Higgs does not have a Yukawa coupling to electrons this limit is completely relaxed.~We can turn this around and say that should CP violation be observed in $h\to4\ell$ due to these couplings, then we also indirectly discover a second BSM effect, e.g.~that the Higgs does not have SM couplings to first generation fermions.~For other frameworks in which this is realized and the EDM constraint evaded see~\cite{McKeen:2012av}.~Irrespective of the EDM constraint, the sign of the $h\gamma\gamma$ coupling is not constrained in general.~We thus conclude that an independent measurement of $A^{\gamma\gamma}_2$ and $A^{\gamma\gamma}_3$ at the LHC is desirable.

\emph{\underline{Couplings to $ZZ$}}:~CMS and ATLAS have tested the hypothesis of a pure scalar coupling $A_1^{ZZ}$ versus pure pseudo scalar coupling $A_3^{ZZ}$ using the differential distributions of leptons in the $4\ell$ channel, each excluding a pure pseudo-scalar at about~$3\sigma$~\cite{htoAA}.~CMS has also put constraints on CP odd/even mixtures and finds a CP odd component as large as $\sim40\%$ is still allowed~\cite{Chatrchyan:2012jja,Chatrchyan:2013mxa,CMS-PAS-HIG-14-014, Khachatryan:2014kca}.~Assuming that $A_1^{ZZ}$ is indeed highly dominant, as expected from the dimensionality of the operators and EWSB in the SM, this coupling is constrained from the total rate of $h \to ZZ \to X$ to be around $1.43\pm0.4$ (ATLAS) and $0.92\pm0.28$ (CMS) times the SM value~\cite{htoAA,Aad:2013wqa}.~In our work we will simply fix it to the tree level SM value of $A_1^{ZZ} = 2$ and treat it as a background to the other couplings in~\eref{Aall}.

\emph{\underline{Couplings to $Z\gamma$}}:~The coupling of the Higgs to a photon and a $Z$ is currently poorly constrained from the direct $h\to Z\gamma$ decay, and is expected to remain so in the near future.~The current bound from CMS on the relevant signal strengths is 13.5 times the SM expectation and thus not yet sensitive to the SM values of $A^{Z\gamma}_2\sim 0.014$ and $A^{Z\gamma}_3\sim0$.~The projected precision for CMS on the signal strength into $Z\gamma$, which is proportional to $|A^{Z\gamma}_2|^2 + |A^{Z\gamma}_3|^2$, is 62\% with 300 fb$^{-1}$ and about 20-24\% with 3000 fb$^{-1}$~\cite{htoAA}, (The ATLAS current projections are worse by a factor of two).~These correspond to a precision of 41\% and 10-12\% on the measured effective couplings.~Any additional way to constrain the $hZ\gamma$ couplings is thus highly desirable.~CMS has also already begun incorporating these couplings into their $h\to4\ell$ analysis~\cite{CMS-PAS-HIG-14-014, Khachatryan:2014kca}, but again the sensitivity is still weak.

To summarize, the couplings of the Higgs to neutral electroweak gauge bosons are partially constrained by current LHC measurements and EDM limits.~However, it is worth emphasizing what we \emph{don't} know.~We do not know the sign of the coupling to photons nor do we have a model independent limit on its CP phase.~We also do not know the magnitude or CP structure of the Higgs couplings to $Z\gamma$.~In this work we will show that a $h\to 4\ell$ analysis can shed light on these interactions during LHC running if it is optimized to do so.  

\section{Current Cuts and Lepton pairings}

The cuts used in current LHC analyses of the four lepton channel were set at a time when the Higgs was not yet discovered.~The goal of these cuts was to enhance the SM $ZZ$ signal over the non-Higgs backgrounds.~For CMS these cuts are approximated by $p_{T\ell} > 20, 10, 7, 7~$GeV for lepton $p_T$ ordering, $|\eta_\ell| < 2.4$ for the lepton rapidity, and $40$~GeV $\leq M_1$ and $12$~GeV $\leq M_2$ for the reconstructed masses of same-flavor opposite-sign lepton pairs.~The CMS prescription for choosing the pairs is to impose $M_1 > M_2$ and to take $M_1$ to be the reconstructed invariant mass for a particle and anti-particle pair which is closest to the $Z$ mass.~This pairing prescription will play an interesting role below.

Our goal in this work is to study how much the sensitivity of this channel to Higgs couplings to $Z\gamma$ and $\gamma\gamma$ can be enhanced by relaxing the standard cuts.~We note however that, due to pairing effects, the $h\to4\ell$ channel is already sensitive to the $h\gamma\gamma$ couplings even with the standard cuts~\cite{Chen:2014gka}.~Naively, one might find this surprising since these cuts would appear to be very efficient at \emph{removing} events in which a lepton pair originated from an off-shell photon since the invariant mass of such a pair would tend to be low.~We could expect that the efficiency for $h\to4\ell$ via $\gamma\gamma$ would thus be particularly low.~As we discuss more below, this turns out not to be the case in the $4e$ and $4\mu$ final states due to `wrong' pairing of leptons which is a consequence of the indistinguishable nature of the final state (same sign) fermions.

Though we use all the observables available in $h\to4\ell$ in our analysis~\cite{Chen:2012jy,Chen:2013ejz,Chen:2014pia,Chen:2014gka}, we can get a good qualitative picture and simplify the discussion by focusing on the lepton pair invariant masses $M_1$ and $M_2$ which alone are already strongly discriminating variables ($M_2$ in particular~\cite{Isidori:2013cla,Grinstein:2013vsa,Gonzalez-Alonso:2014rla}).~In~\fref{m1m2} we show the $M_1$-$M_2$ distribution for several signal operators in~\eref{dim5lag}.~The top panels show the distribution for pure $A_2^{ZZ}$ events, while the middle ones show $A_2^{Z\gamma}$, and the bottom ones show $A_2^{\gamma\gamma}$.~The distributions for the $A_1^{ZZ}$ `background' are very similar to $A_2^{ZZ}$ and thus not shown.~Plots on left show the $2e2\mu$ channel and those on the right show $4e/4\mu$.~In all plots, except for the bottom right the distributions are highly peaked in the region one would expect, where $M_1$ and $M_2$ are near the respective on-shell masses of the $Z$ and photon.~However, in the case of a di-photon mediated amplitude in the $4e/4\mu$ channel (bottom right plot), the spectral peak near $M_{1,2}=0$ is removed and events are instead spread in the bulk of the $M_1$-$M_2$ plane.~As a result the efficiency in the $h\to\gamma\gamma\to4e/4\mu$ channel is much higher than the corresponding $2e2\mu$ channel.~How can we understand the difference between this case and the others seen in~\fref{m1m2}?
\begin{figure}
\includegraphics[width=.23\textwidth]{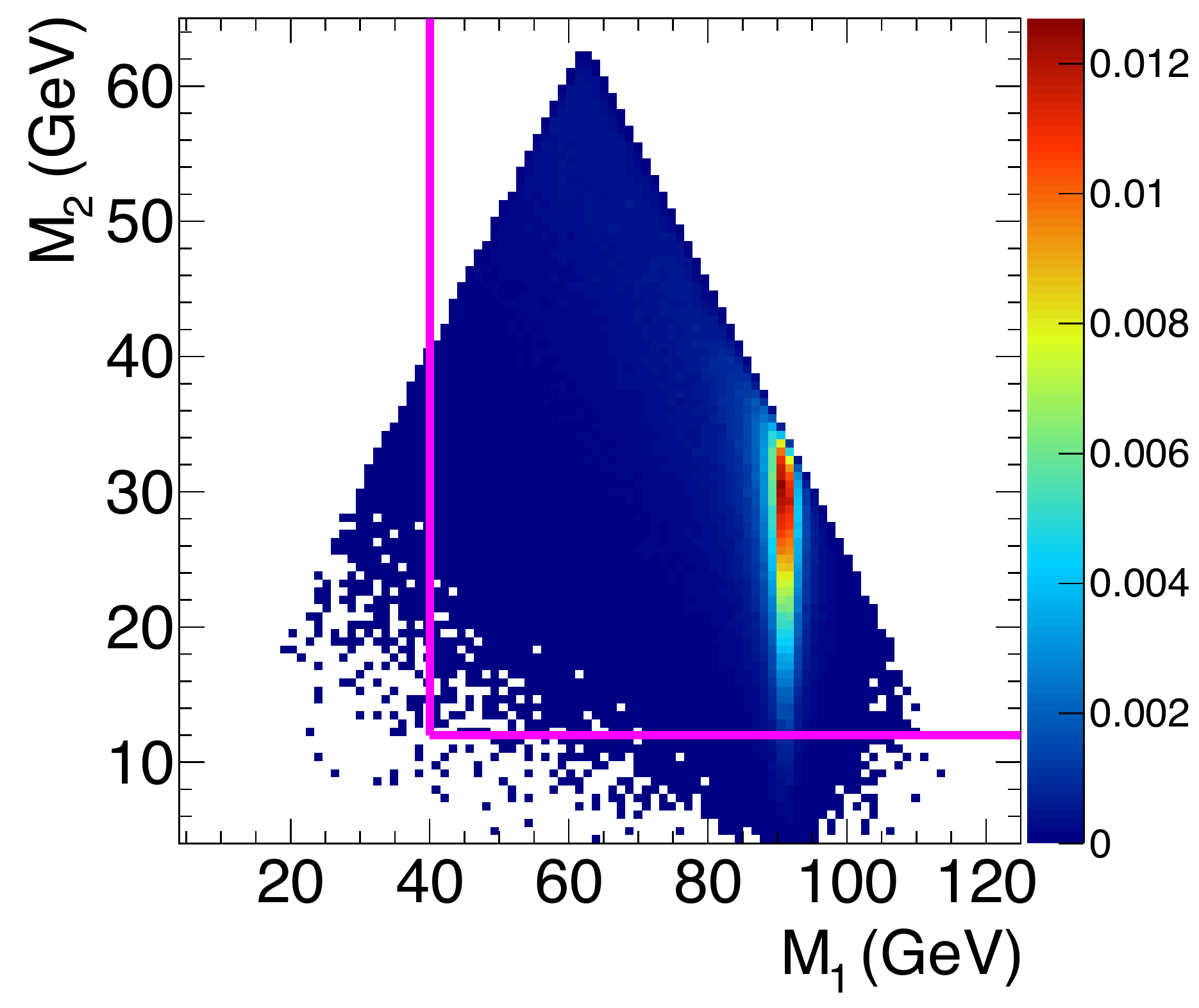}
\includegraphics[width=.23\textwidth]{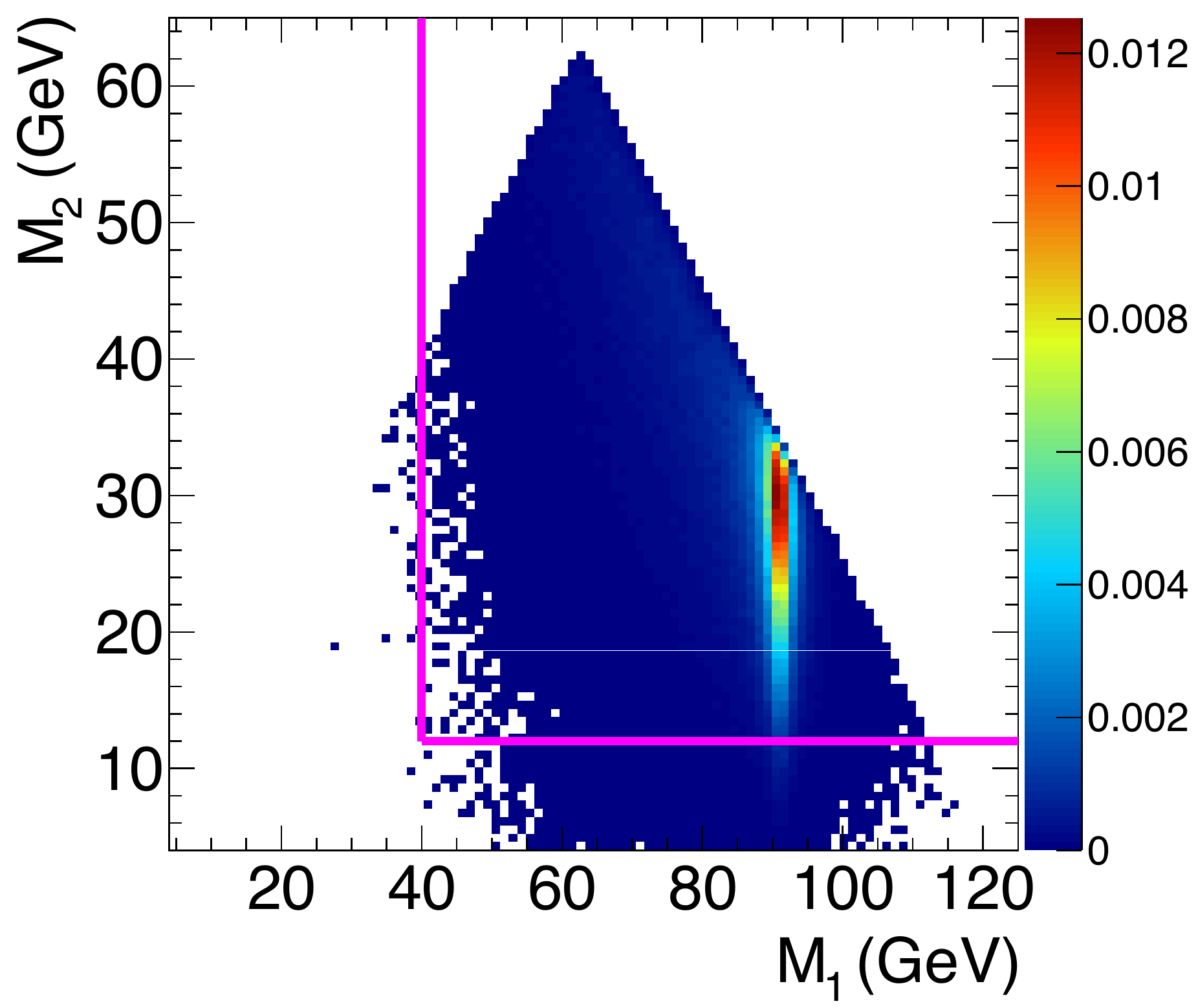}\\
\includegraphics[width=.23\textwidth]{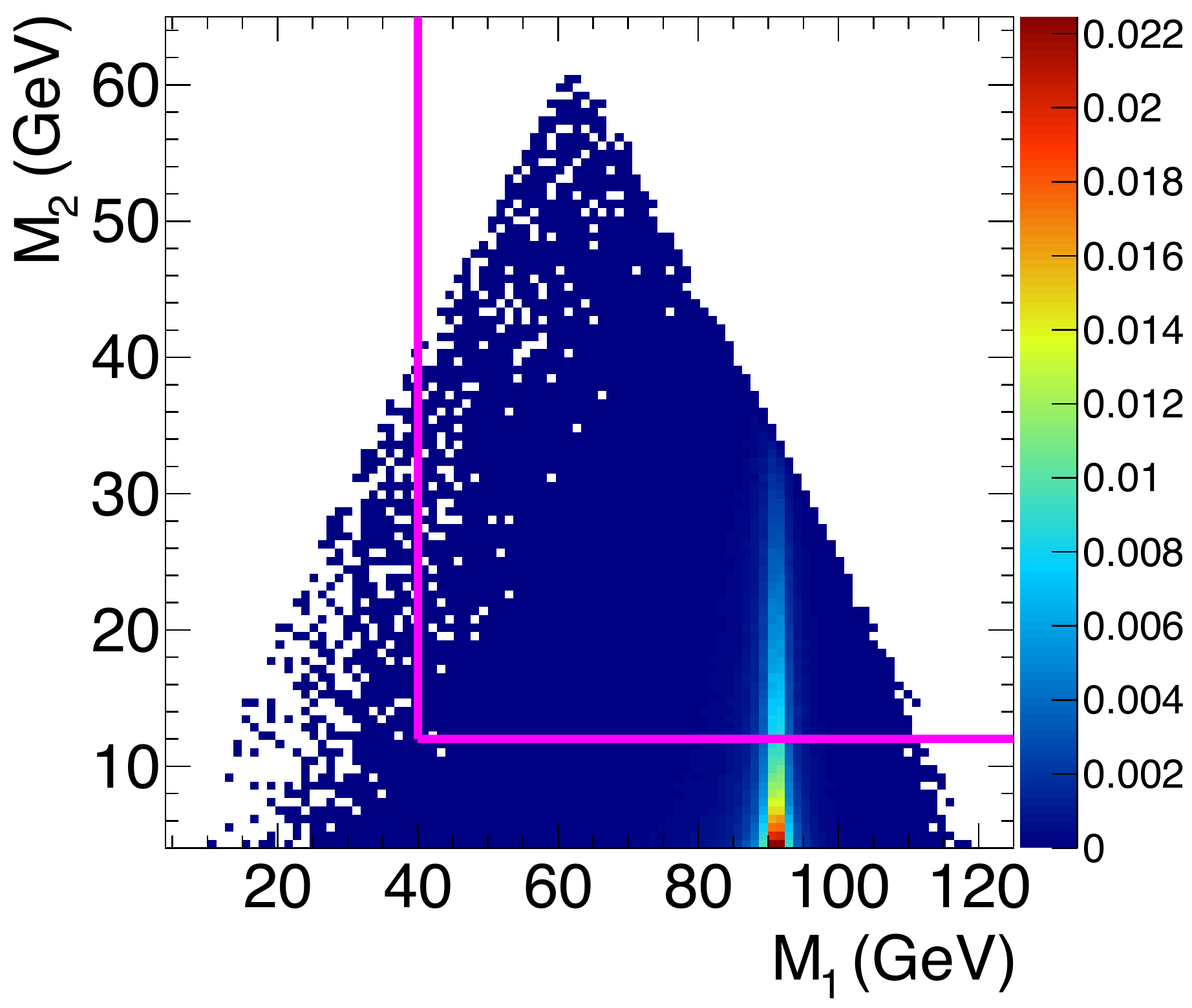}
\includegraphics[width=.23\textwidth]{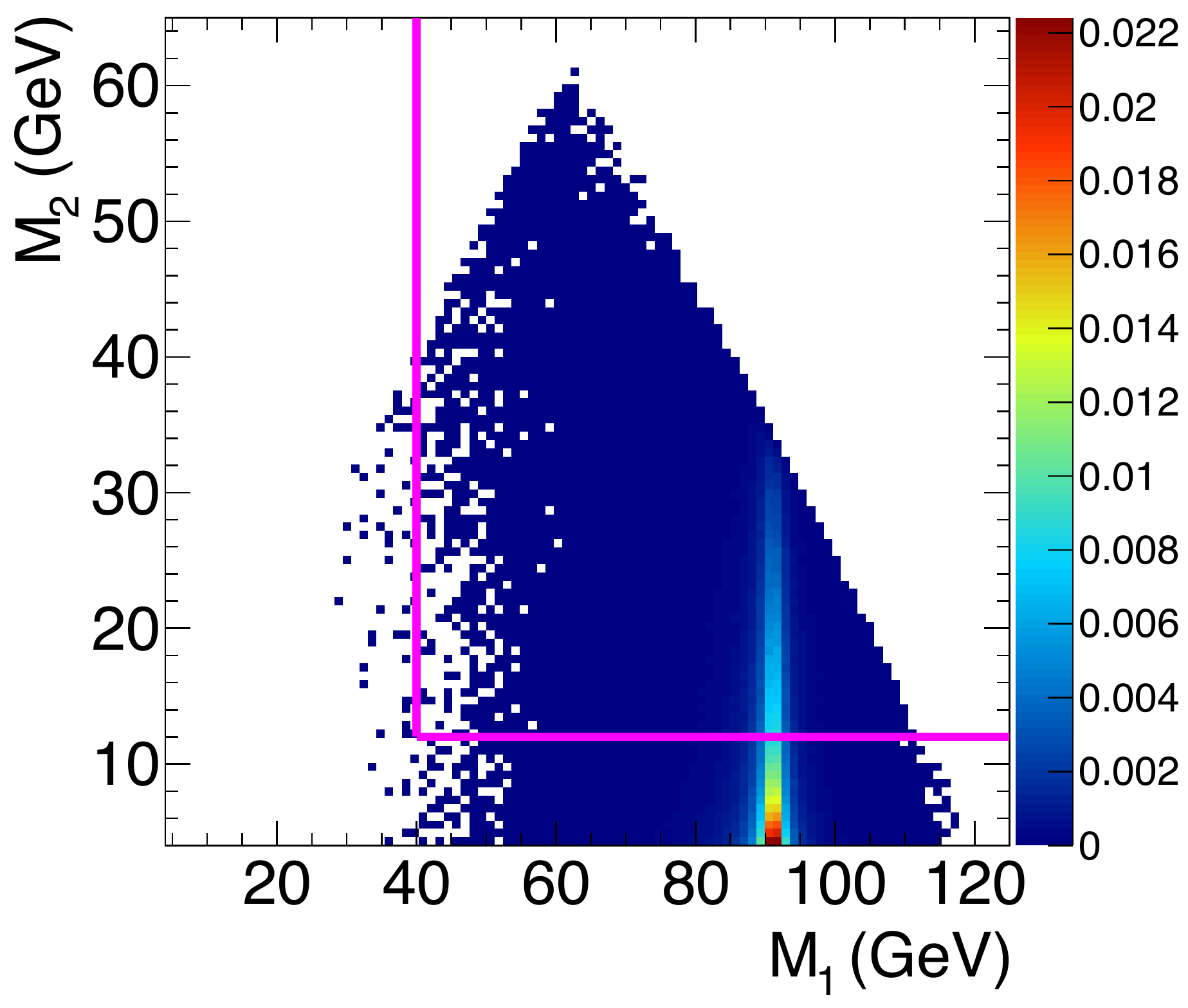}\\
\includegraphics[width=.23\textwidth]{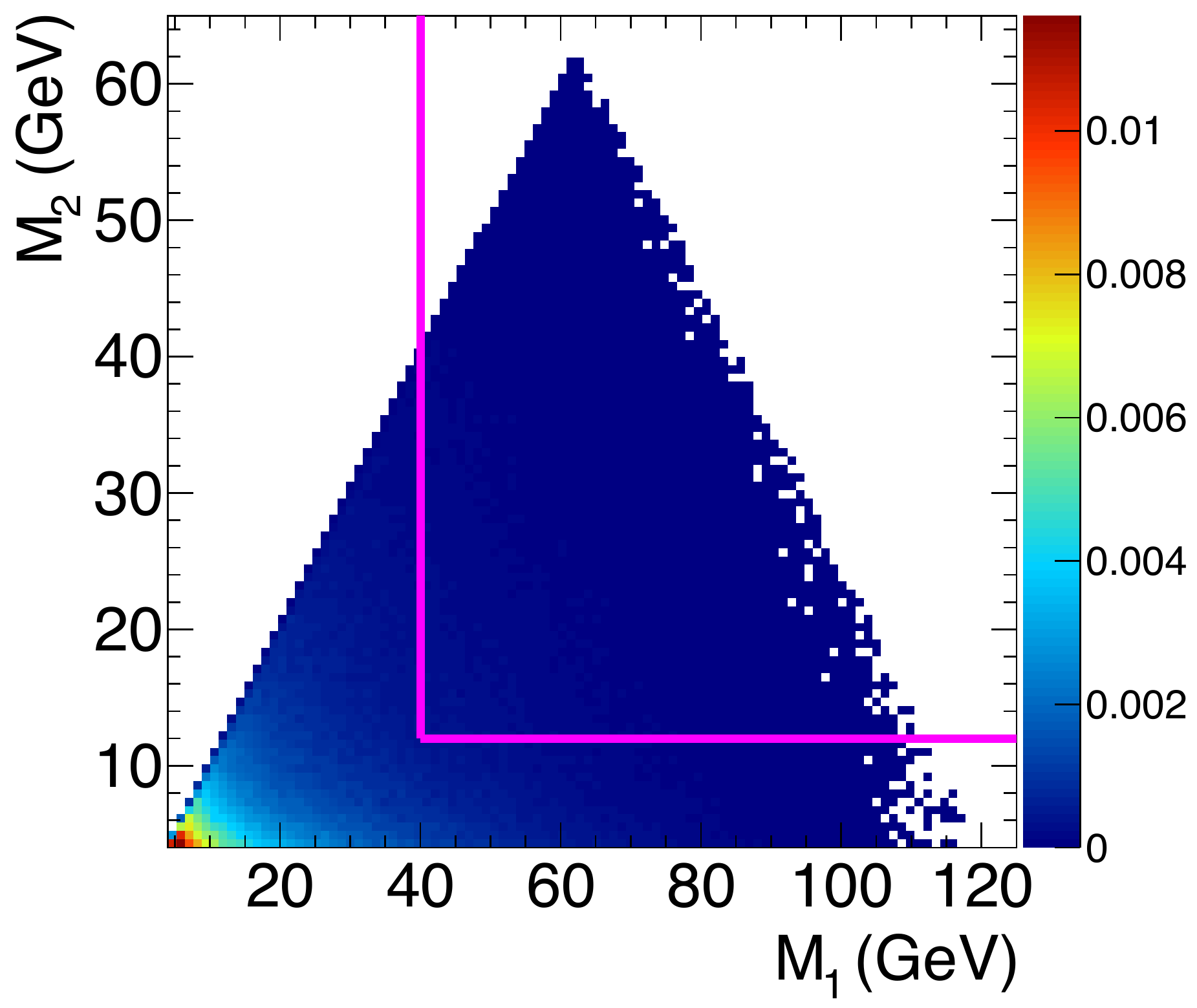}
\includegraphics[width=.23\textwidth]{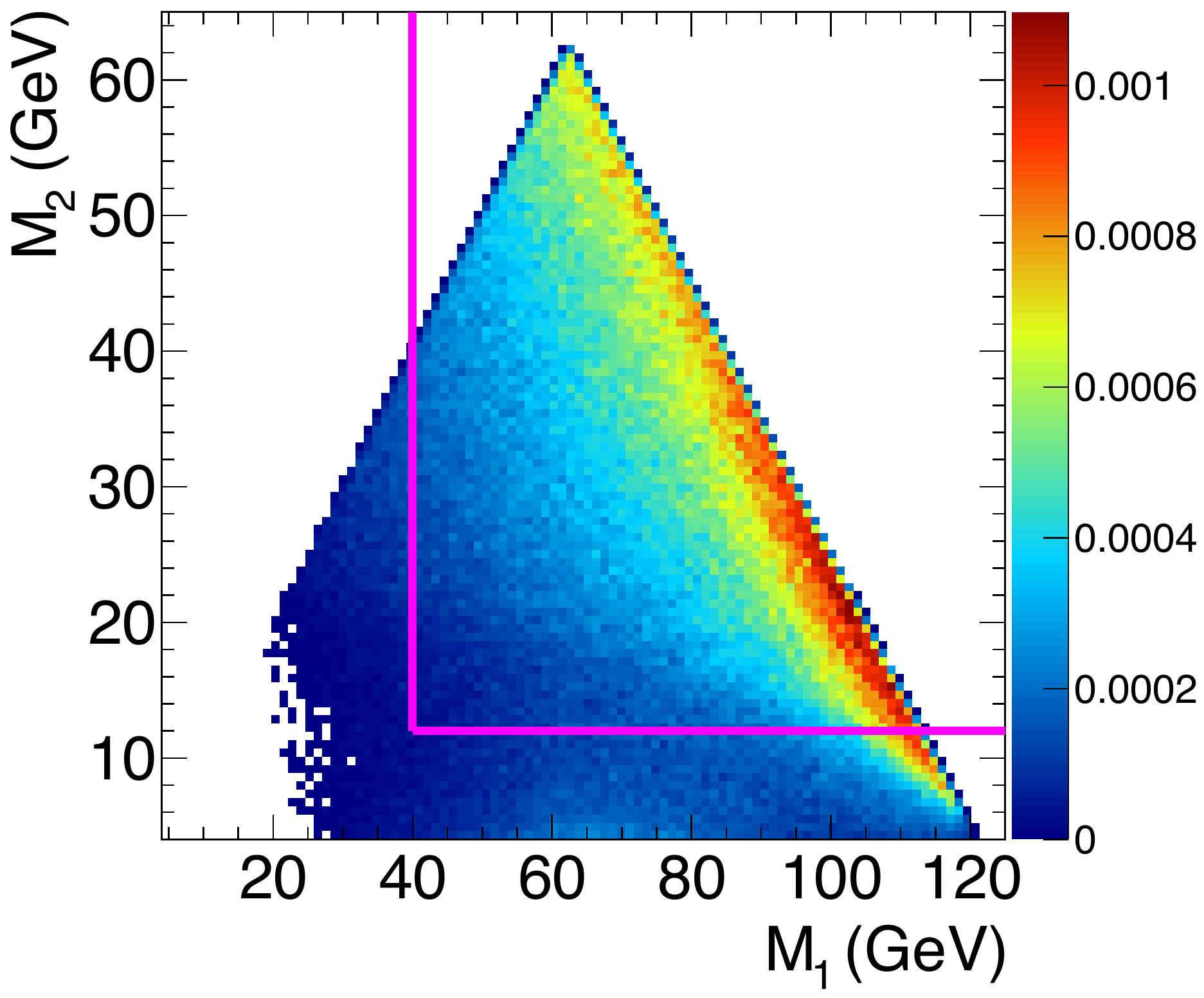}
\caption{{\bf Top:}~$M_1-M_2$ doubly differential distribution assuming only the $A_2^{ZZ}$ operator defined in~\eref{dim5lag} is `turned on' for the $2e2\mu$ final state (left) and the $4e$ final state (right).~{\bf Middle:}~Same as top figures, but now for $A_2^{Z\gamma}$ couplings.~{\bf Bottom:}~Same as top figures, but now for the $A_2^{\gamma\gamma}$ couplings.~For all distributions standard CMS lepton pairings are applied (see text) and the pink lines indicate the $M_1 > 40$~GeV and $M_2 > 12$~GeV cuts used by CMS~\cite{Khachatryan:2014kca}.~``Wrong pairing'' effects are important in the bottom right distribution and discussed more in text.}
\label{fig:m1m2}
\end{figure}

For the $2e2\mu$ final state, $M_1$ and $M_2$ are formed from $e^+e^-$ and $\mu^+\mu^-$ (or vice versa).~The $\gamma\gamma$ component of the $h\to4\ell$ amplitude has no ambiguity in this case and thus each pair does originate from an off-shell photon.~Therefore, the di-photon amplitude does indeed peak at low values of $M_1$ and $M_2$ and the standard cuts effectively remove this component.~For the $4e$ and $4\mu$ final states, the identical final states introduces an additional, but equally valid, pairing obtained by swapping the electrons (or muons) or positrons (or anti-muons).~The prescription used to resolve this ambiguity, picking $M_1$ to be closer to the $Z$ mass, implicitly assumes that there is a nearly on-shell $Z$ in the process.~However, this assumption does not hold for the signal amplitudes that are mediated by two off-shell photons.~As a result, for almost all `$\gamma\gamma$ events' the lepton pair that is chosen to make up $M_1$ does not originate from the same photon, but rather from two different photons that are back-to-back in the Higgs frame (hence maximizing the lepton pair invariant mass).~A heuristic sketch of this `wrong pairing' effect is shown in~\fref{cartoon}.~It should be noted however that due to quantum interference no event is purely $ZZ$, $Z\gamma$, or $\gamma\gamma$.~In addition, even restricting to $\gamma\gamma$ amplitudes, there is a small interference among the different pairing choices (see~\cite{Chen:2013ejz}) in $4e$ and $4\mu$, though this interference effect is small over most of the phase space and the heuristic argument above goes through.~A similar argument can be applied to the CP odd $A^i_3$ couplings since their $M_1-M_2$ distributions are similar (but again not identical) to those for the CP even couplings.
\begin{figure}[h]
\begin{center}
\includegraphics[width=.5\textwidth]{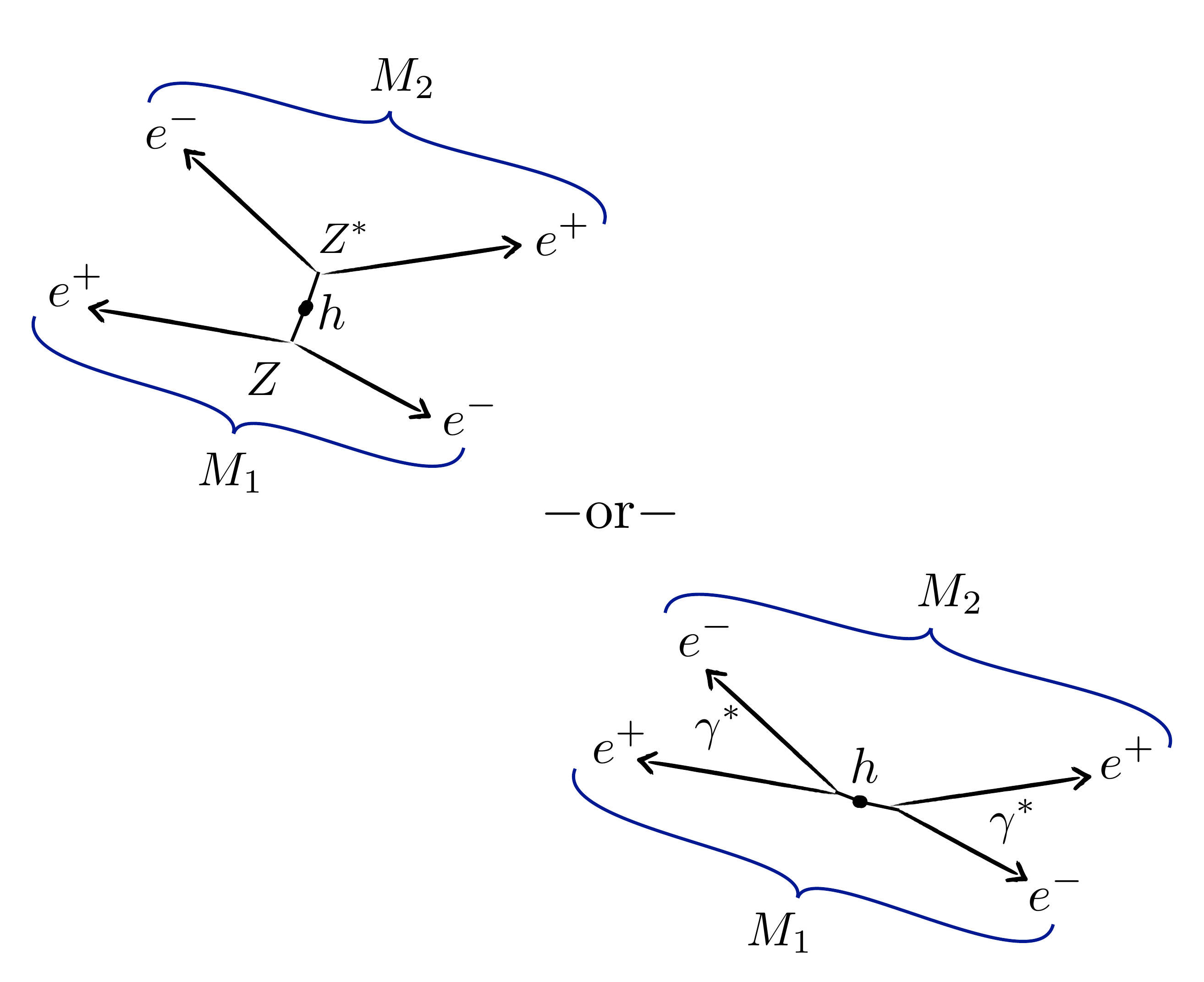}
\end{center}
\caption{Heuristic sketch of the difference in lepton pairings between $ZZ$ events and $\gamma\gamma$ events.~The wrong lepton pairing in the $\gamma\gamma$ case significantly increases the acceptance of such events in the $4e$ and $4\mu$ channels.}
\label{fig:cartoon}
\end{figure}

This ``wrong pairing'' effect and the increased efficiency is a major factor in the ability of the current analyses (with more data) to probe the $h\gamma\gamma$ coupling~\cite{Chen:2014gka} and also implies that the sensitivity is driven by the $4e$ and $4\mu$ channels.~This can be seen explicitly in~\fref{EMvsEE} where we show sensitivity curves for the `average error' $\sigma(A_2^{\gamma\gamma})$ on the extracted value (as defined in~\cite{Chen:2014gka}) of $A_2^{\gamma\gamma}$ as a function of the number of events.~In these curves we have applied the current CMS-like cuts and fit to a `true' point of $\vec{A} = (0,0,0,0,0,0)$.~We indicate by the green dashed line the magnitude for the SM value of $|A_2^{\gamma\gamma}| = 0.008$~\cite{Low:2012rj}.~We see clearly that the `accidentally' high acceptance for the $\gamma\gamma$ component in $4e$ (or $4\mu$) leads to a significantly stronger sensitivity to the $h\gamma\gamma$ couplings than in the $2e2\mu$ channel.~As expected from~\fref{m1m2} we find that the sensitivity to $ZZ$ and $Z\gamma$ is similar in the two channels and thus we do not show the curves.~Since the acceptance is largely determined by the $M_1-M_2$ distributions we also do not show the curves for the CP odd coupling $A_3^{\gamma\gamma}$ which show a similar (but not identical) behavior to the curves in~\fref{EMvsEE} for the CP even coupling.
\begin{figure}[tbh]
\begin{center}
\includegraphics[width=.45\textwidth]{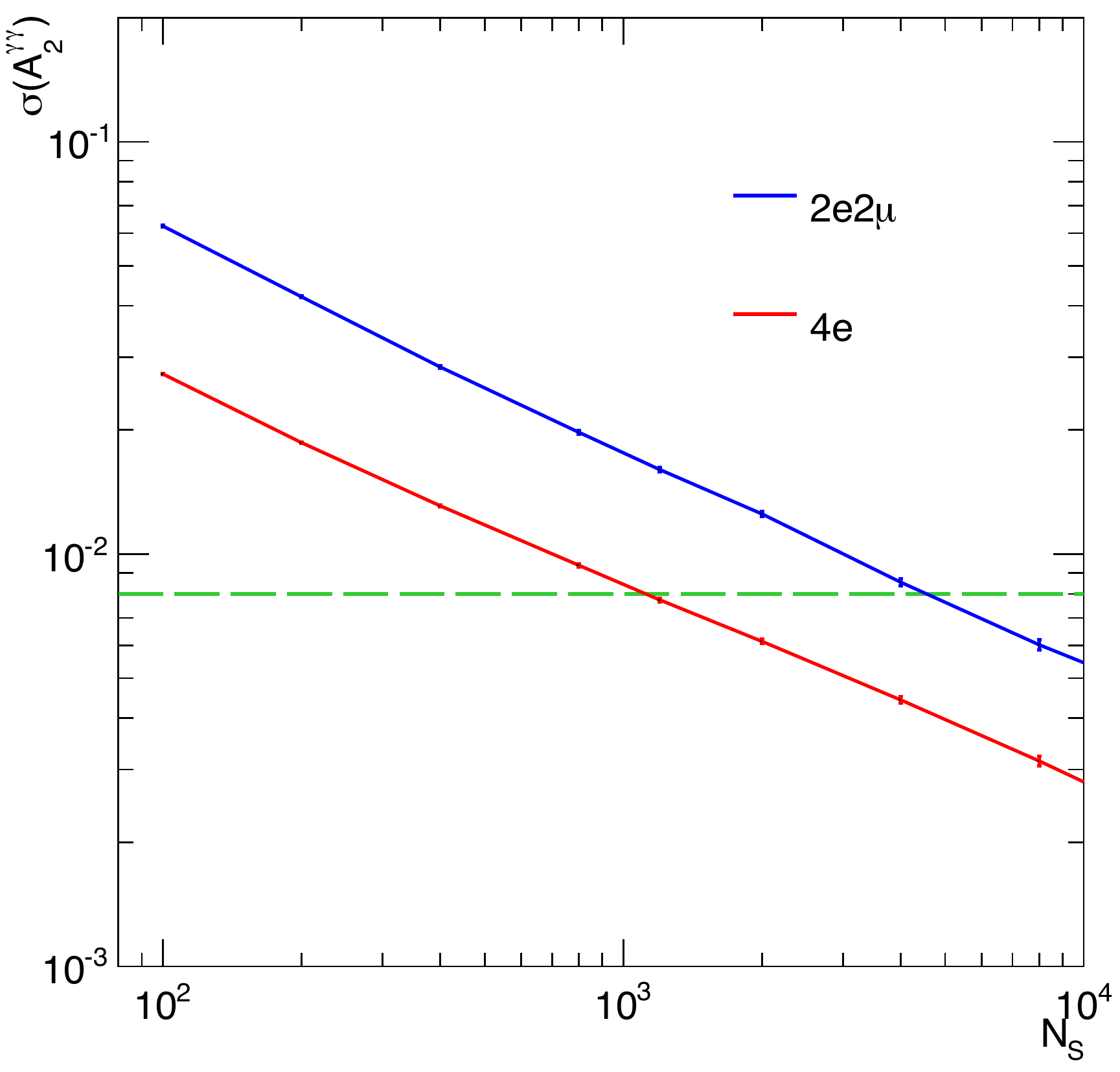}
\end{center}
\caption{Comparison of sensitivity to $A_2^{\gamma\gamma}$ as a function of number of events in $2e2\mu$ (blue) and $4e$ (red) channels for CMS-like cuts.~Here we have fit to a pure signal data sample and a true point of $\vec{A} = (0,0,0,0,0,0)$ while floating all couplings in~\eref{Aall} simultaneously.~We indicate by the green dashed line the magnitude for SM value of $|A_2^{\gamma\gamma}| = 0.008$~\cite{Low:2012rj}.~The sensitivity is quantified by the effective $\sigma(A_2^{\gamma\gamma})$, or average error as defined in~\cite{Chen:2014gka}.}
\label{fig:EMvsEE}
\end{figure}

These considerations also lead us to suspect that there is room for the $h\to4\ell$ analysis to be optimized further.~For example, we could have purposefully made the `wrong pairing' of leptons even in the $2e2\mu$ channel leading to a similar distribution to that seen in the bottom right plot in~\fref{m1m2} for the $4e$ channel.~This of course leads to an enhanced acceptance giving a sensitivity comparable to the $4e$ and $4\mu$ channels.~Of course if the entire phase space is considered then all pairing choices are equivalent (we are assuming massless leptons) which implies the enhancement from the `wrong pairing' can also be achieved by keeping the current pairing convention but relaxing the cuts in the $M_1-M_2$ plane.~Doing so, as expected, will also help to enhance the sensitivity to the $Z\gamma$ component (see~\fref{m1m2}).~As we shall see, the enhancement is such that the golden channel may even become competitive/complimentary with the Higgs decay to on-shell $Z\gamma$ at the LHC.~The $ZZ$ component on the other hand is minimally affected as expected from the distributions in~\fref{m1m2}.~Thus, as discussed in~\cite{Chen:2014gka}, because the distributions of the higher dimensional $ZZ$ couplings are similar to the tree level $ZZ$ coupling, the $h\to4\ell$ channel is not as strongly sensitive to the $ZZ$ couplings in~\eref{dim5lag} regardless of the cuts or lepton pairings used.

\section{Alternative Cuts and Pairings}

The previous discussion implies that choosing alternative pairings or loosening the cuts on the lepton pair invariant masses, $M_1$ and $M_2$, can enhance the sensitivity to the $h\gamma\gamma$ as well as the $hZ\gamma$ couplings.~As a first demonstration of this we consider various lepton pairings and cuts defined in~\tref{cuts}.~First we have the standard CMS-like cuts for which we have considered two cases, a `CMS-tight' and a `CMS-loose' which differ in the lepton $p_T$ requirements.~We also consider two alternative lepton pairings.~The first is nicknamed `Opposite' and takes the opposite lepton pairing with respect to the CMS choice with the pairs carrying opposite charge, but not necessarily same flavor.~For example in the $2e2\mu$ channel the parings would be $e^-\mu^+$ and $e^+\mu^-$ while for $4e$ and $4\mu$ it would simply be the `other' possible opposite charge lepton pairing that is not the CMS one.~The second alternative pairing we consider is nicknamed `Same' and takes the same sign leptons in each pair.~We also consider the case `Combined' where all three pairings are combined if either the CMS-loose, Opposite, or Same cuts/pairings are satisfied.~We then consider `Relaxed' cuts where we take the CMS pairings and require simply $M_{1,2} > 4$~GeV along with the lepton $p_T$ and $\eta$ cuts.
\begin{table*}[tb]
\begin{center}
\includegraphics[width=.98\textwidth]{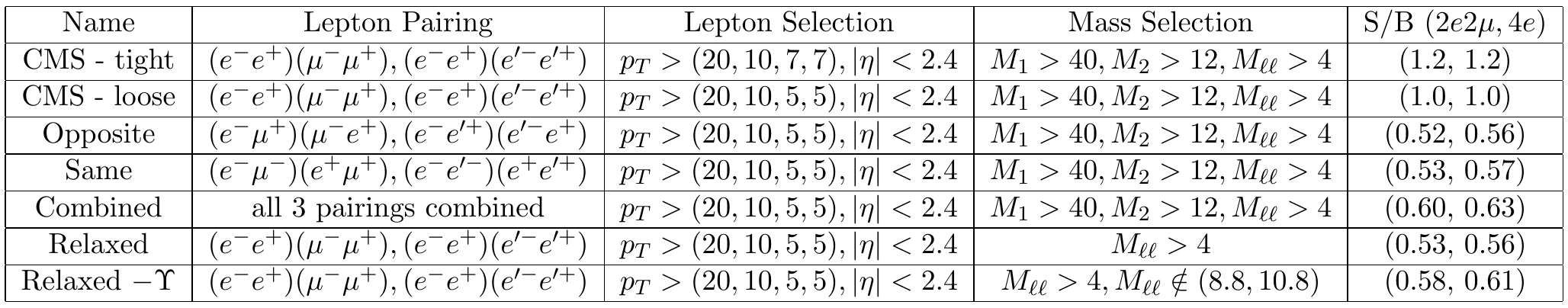}
\end{center}
\caption{The various cuts and lepton pairings which are explored for a four lepton invariant mass range of $115-135$~GeV.~The first column gives the name of the cuts/pairings.~The second column indicates the paring chosen for the case of $2e2\mu$ and $4e$.~The third column indicates the cuts on $M_1$ and $M_2$ as well as any lepton pair $M_{\ell\ell}$.~Finally the last column gives the signal to background ratio for the $2e2\mu$ and $4e$ final states (see text for further information).}
\label{tab:cuts}
\end{table*}

However, relaxing the cuts on $M_1$ and $M_2$ this much introduces contamination from $\Upsilon$ decays.~To avoid the $\Upsilon$ we also consider `Relaxed - $\Upsilon$' cuts where again we require $M_{1,2} > 4$~GeV, but remove events with $8.8$~GeV $< M_{1,2} < 10.8$~GeV.~This will of course reduce the efficiency in the $Z\gamma$ and more so $\gamma\gamma$ components, though not dramatically ($\sim 3-5\%$).~Furthermore, by always requiring $M_{1,2} > 4$~GeV, we also avoid other QCD resonances and large $Z-\gamma$ mixing effects~\cite{Gonzalez-Alonso:2014rla} which distort the spectrum in the very low $M_{1,2}$ region.~However, these $\Upsilon$ effects can be computed~\cite{Gonzalez-Alonso:2014rla} and in principle incorporated into the present framework to enhance the sensitivity further still, but we do not explore that here.

We note that there is no clear roadblock to relaxing the cuts even further, going below $M_{1,2}$ of 4~GeV, particularly in the 4$\mu$ channel.~As an example we refer to a CMS search for the decay of the Higgs to two `dark photons'~\cite{Chatrchyan:2012cg} in which the search region for $M_{1,2}$ is between about 0.25 and 3.5 GeV.~The QCD resonances in this region were accounted for using a data driven method.~In fact it is interesting to consider recasting this search in order to place a constraint on the Higgs couplings to photons, but we leave this for a future study.

\vspace*{-.3cm}
\subsection{Effects of Cuts on Sensitivity}

For each of these cuts and lepton pairings in~\tref{cuts} we examine sensitivity curves of $\sigma(A_n^i)$ as a function of $N/\epsilon$ where $N$ is the number of events and $\epsilon$ is the selection efficiency for a given set of cuts.~The results are shown in~\fref{cutscurves} where again we have fit to a `true' point of $\vec{A} = (0,0,0,0,0,0)$ on a pure signal data sample.~On the left we show results for the sensitivity to the $A_2^{Z\gamma}$ couplings while on the right we show the sensitivity to $A_2^{\gamma\gamma}$.~Note that all couplings in~\eref{Aall} are floated simultaneously and no assumptions about relations between the couplings are made.
\begin{figure*}[tbh]
\begin{center}
\includegraphics[width=.45\textwidth]{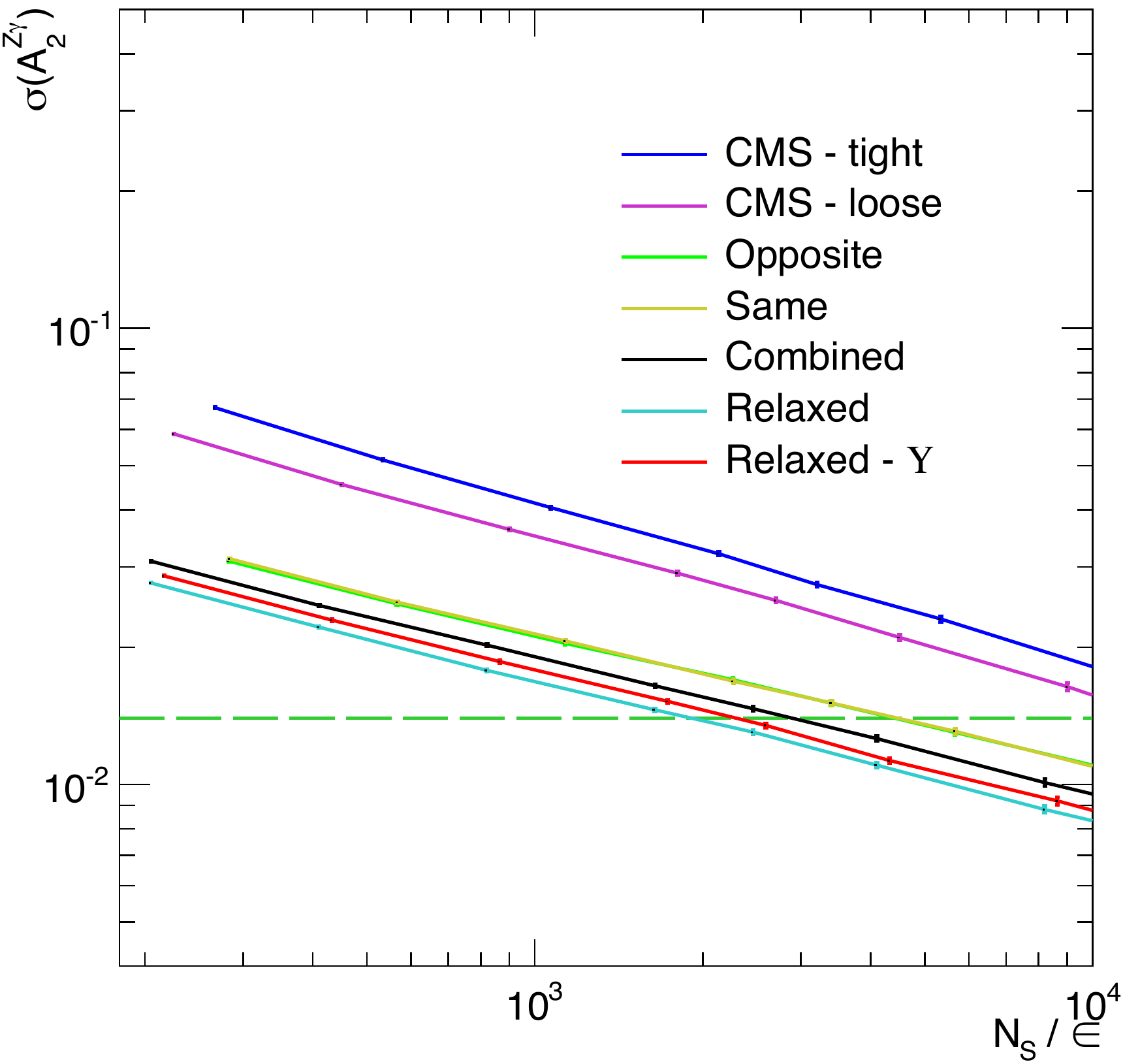}
\includegraphics[width=.45\textwidth]{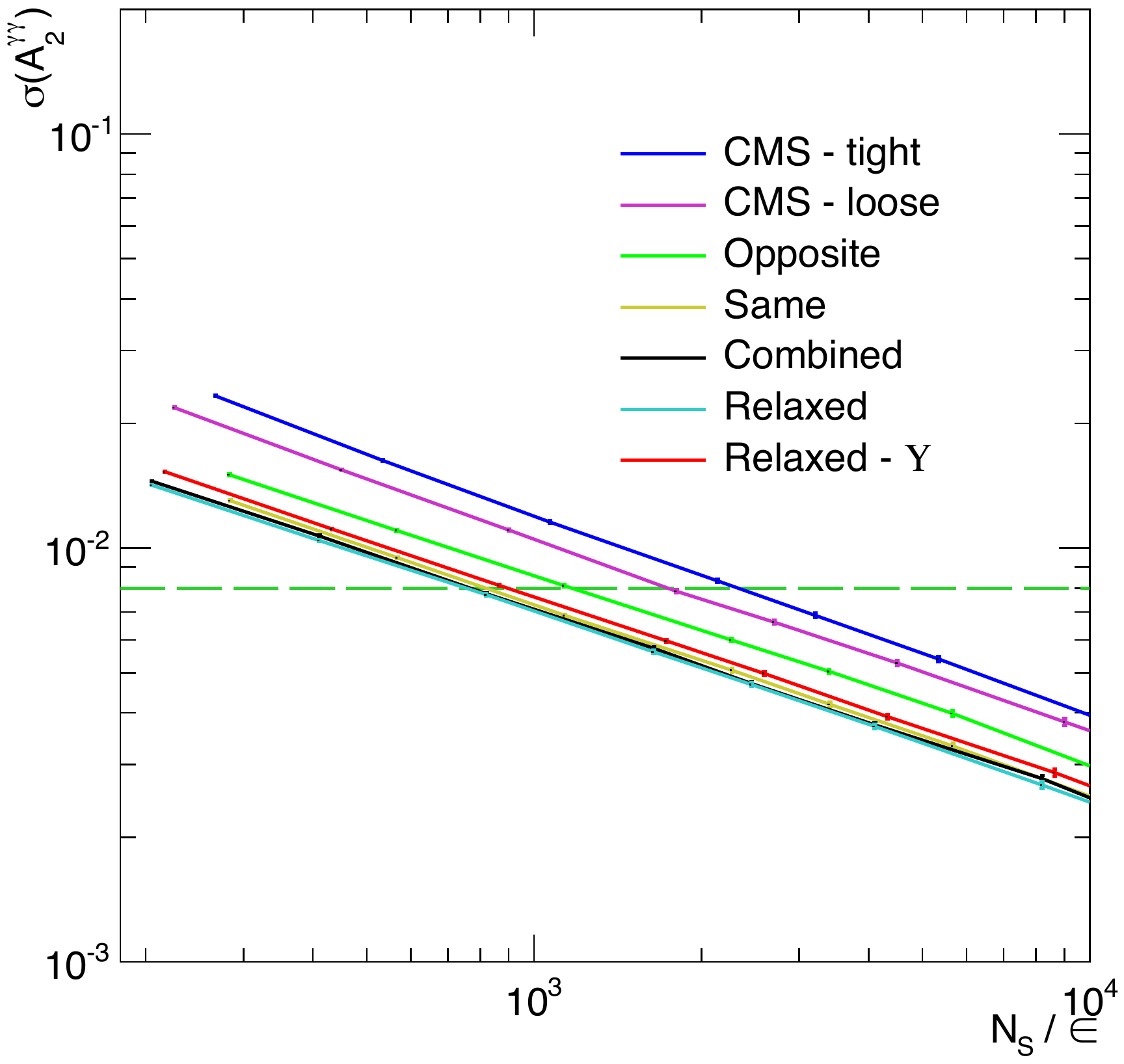}
\end{center}
\caption{Sensitivity curves for $A_2^{Z\gamma}$ (left) and $A_2^{\gamma\gamma}$ (right) as function of number of events divided by efficiency for the combined $2e2\mu, 4e, 4\mu$ channels for the sets of cuts and lepton pairings described in~\tref{cuts}.~We have fit to a true point of $\vec{A} = (0,0,0,0,0,0)$ on a pure signal data sample and floated all couplings in~\eref{Aall} simultaneously.~The green dashed lines indicate the magnitude for the SM value of $|A_2^{Z\gamma}| = 0.014$ and $|A_2^{\gamma\gamma|}| = 0.008$~\cite{Low:2012rj}.}
\label{fig:cutscurves}
\end{figure*}

\emph{\underline{Effects of cuts on $hZ\gamma$}}:~We first turn our attention to the $A_2^{Z\gamma}$ coupling, shown in the left of~\fref{cutscurves}.~The first thing to notice is that the sensitivity is drastically improved \emph{wrt} the current CMS-like cuts (blue and pink) using any of the other cuts or lepton pairings in~\tref{cuts}.~In particular, whereas with CMS like cuts $\gtrsim 20,000$ events (including efficiency) are needed to probe values of order the SM value, now we see in the most optimistic case of Relaxed cuts (turqoise), only $\lesssim 2000$ events may be needed.~This in principle makes this an LHC question as opposed to certainly needing a $100$~TeV collider (or some other future machine) to probe SM values of these couplings in $h\to4\ell$.

We see also in the Relaxed $-\Upsilon$ cuts (red) that removing events around the $\Upsilon$ mass does mildly affect the sensitivity to where now $\gtrsim 2000$ events are needed, thus perhaps allowing for further optimization by including this region.~Note that the two alternate pairings Opposite (green) and Same (gold) perform equally well and much better than the pairings in CMS-cuts.~This is because both of these pairings lead to a similar $M_1-M_2$ spectrum which no longer is peaked at $M_2 \sim 0$, thus greatly enhancing the efficiency.~Note also that Relaxed cuts perform noticeably better than the Combined cuts (black) showing that combining all three pairings is similar, but not equivalent to keeping the standard pairing and lowering the $M_{1,2}$ cuts.~Finally we can also see by comparing the CMS-tight and CMS-loose that relaxing the lepton $p_T$ does improve the sensitivity noticeably.~Similarly qualitative features are seen for the CP odd $A_3^{Z\gamma}$ so we do not show it separately.

\emph{\underline{Effects of cuts on $h\gamma\gamma$}}:~Turning now to the $A_2^{\gamma\gamma}$ coupling, shown in the right of~\fref{cutscurves}, we see a number of quantitatively and qualitatively different features than for $A_2^{Z\gamma}$.~The first of course is the stronger sensitivity in general to the $h\gamma\gamma$ coupling, though the alternative cuts and pairings give a less drastic improvement over the CMS like cuts than seen for $Z\gamma$.~In particular we see that although standard CMS cuts are sufficient to begin probing SM values of these couplings with $\sim 2000-3000$ events, with the Relaxed cuts this is reduced down to $\sim 700-800$ events (again including efficiency).~Removing the events around the $\Upsilon$ mass reduces the sensitivity somewhat requiring $\sim 900$ to reach the necessary sensitivity.~Interestingly in the case of $\gamma\gamma$ the Same and Opposite pairings do not perform equally and in fact the Same pairing performs nearly as well as the Relaxed cuts.~This is because in the $4e$ and $4\mu$ channels the Opposite pairings leads to a $M_1-M_2$ spectrum which looks like the bottom left plot in~\fref{m1m2}, thus severely degrading the efficiency.~On the other hand the Same pairing gives a spectrum for all three final states which looks like the bottom right plot in~\fref{m1m2} leading to a large acceptance.~Again the CP odd $A_3^{\gamma\gamma}$ coupling shows qualitatively similar behavior so we do not show it separately.

The sensitivity curves seen in~\fref{cutscurves} demonstrate that utilizing alternative cuts or lepton pairings can indeed significantly enhance the sensitivity to the $hZ\gamma$ and $h\gamma\gamma$ couplings.~Of course, as expected, the Relaxed cuts (turquoise) perform best for both $Z\gamma$ and $\gamma\gamma$ since they encompass the largest phase space even though $S/B$ is smaller than for CMS-like cuts (see~\tref{cuts}).~By comparing these cuts to the Relaxed $-\Upsilon$ cuts we can see the effect of cutting out events which fall near the $\Upsilon$ mass.~We see that as expected from the distributions in~\fref{m1m2}, removing events around the $\Upsilon$ mass degrades the sensitivity to the $h\gamma\gamma$ coupling (proportionally) more than to $hZ\gamma$.~Though the reduction in sensitivity is not drastic, one can see that a proper treatment of this region such as done in~\cite{Gonzalez-Alonso:2014rla} can help in optimizing the sensitivity to these couplings.

To summarize, we find that the sensitivity to our signal $hZ\gamma$ and $h\gamma\gamma$ couplings is enhanced by modifying the current analysis cuts.~Not surprisingly, the Relaxed and `Relaxed-$\Upsilon$' cuts show the greatest enhancements.~In what follows we will study the sensitivity of the four lepton channel to all of the couplings in~\eref{Aall}, including the dominant non-Higgs backgrounds, focusing on the Relaxed-$\Upsilon$ and CMS-tight cuts.~Before doing so, we first examine the relative sizes of the various contributions to the $h\to 4\ell$ decay when utilizing the Relaxed-$\Upsilon$ cuts and how these compare to when CMS-tight cuts are used.

\vspace*{-.2cm}
\subsection{The Integrated Magnitudes}
We can gain an intuitive feel of the relative sensitivity to the various couplings by considering the `integrated magnitudes' which were introduced and defined in~\cite{Chen:2014gka}:
\bea
\label{eqn:intmags}
\Pi^{ij}_{nm}=A^i_n A^j_m \times \int \left| \frac{d\Gamma^{ij}_{nm}}{d\mathcal{O}}\right|d\mathcal{O} .
\eea
Here $\mathcal{O}$ is the full set of kinematic observables in the $h\to4\ell$ decay.~Roughly speaking these magnitudes quantify by how much the various Higgs coupling affect the fully differential distribution.~As discussed in~\cite{Chen:2014gka}, these give a better indicator of the size of interference effects than the true total partial widths since these integrated magnitudes contain information not only about the total phase space contribution of each combination of operators, but also about the differences in shape.~It is for this reason that one can have non-zero values even for combinations of operators which lead to CP violation.~The integrated magnitudes are shown in~\fref{absmat} for Relaxed$-\Upsilon$ cuts for the $2e2\mu$ (top) and $4e/4\mu$ (bottom) final states.~To obtain these values we have set $A_1^{ZZ} = 2$ and all other couplings to $A_n^i = 1$ while normalizing to the tree level SM value for the $h\to4\ell$ partial width.~This corresponds to $A_1^{ZZ} = 2$ and all other couplings zero giving unity for the $A_1^{ZZ}\times A_1^{ZZ}$ entry.

The values for $2e2\mu$ in the top of~\fref{absmat} are to be compared to those obtained in~\cite{Chen:2014gka} for CMS-tight cuts.~The crucial thing to notice for $2e2\mu$ is the larger ($\sim 15-60\%$) interference between the $\gamma\gamma$ and $Z\gamma$ couplings with the tree level $A_1^{ZZ}$ coupling (bottom row in tables).~As discussed in~\cite{Chen:2014gka}, since for couplings of order the SM values ($\lesssim \mathcal{O}(10^{-2}-10^{-3})$), the sensitivity is driven by these interference terms, we expect the sensitivity to these couplings to also be enhanced when using the Relaxed$-\Upsilon$ cuts as compared to the standard CMS-like cuts.~Also as discussed in~\cite{Chen:2014gka}, these integrated magnitudes help to qualitatively explain the various shapes seen in the sensitivity curves.~Combined with differences in shapes~\cite{Chen:2014gka} these numbers support the fact that the strongest sensitivity is found for the $Z\gamma$ and $\gamma\gamma$ couplings.~Note also that the tables in~\fref{absmat} and~\cite{Chen:2014gka} can easily be used in any new physics model which predicts values for the various $A_n^i$ to obtain the integrated magnitude and thus a rough estimate on possible contributions to $h\to4\ell$. 
\begin{figure}[tbh]
\includegraphics[width=.45\textwidth]{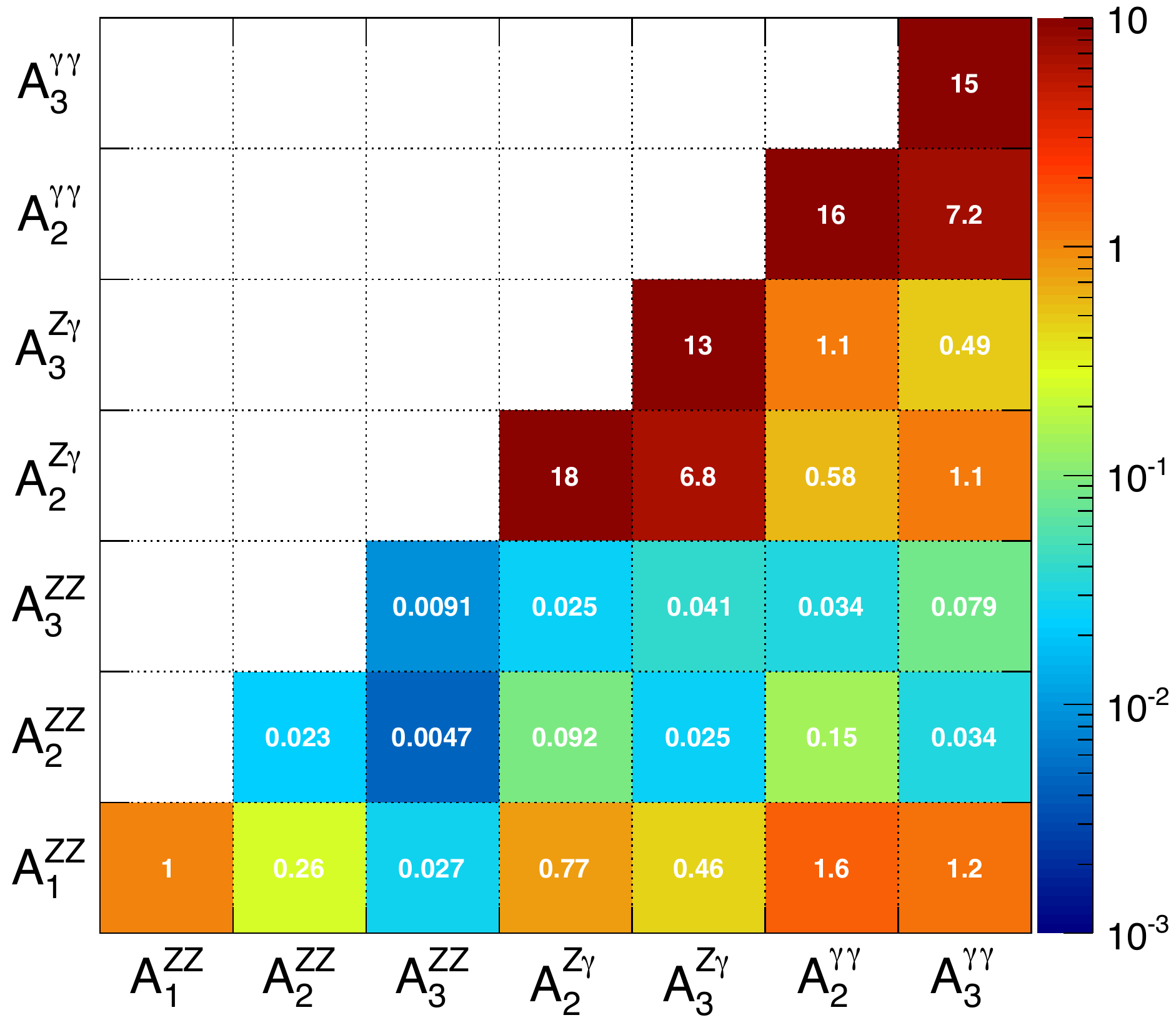}
\includegraphics[width=.45\textwidth]{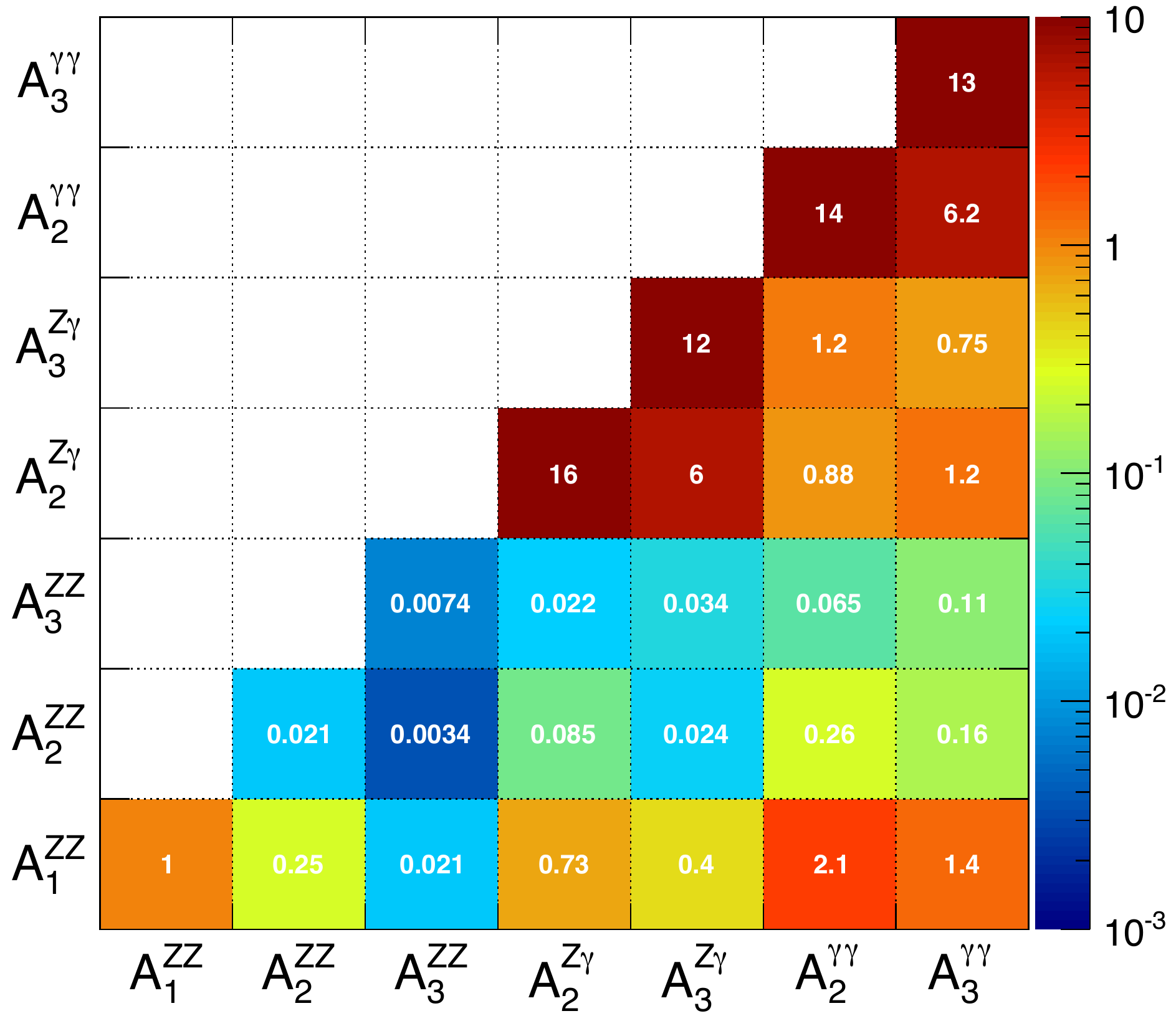}
\caption{{\bf Top:}~The total~\emph{integrated magnitudes}, defined in~\eref{intmags}, corresponding to the pairs of couplings $A^i_n A^{j}_m$ for the $2e2\mu$ final state and Relaxed$-\Upsilon$ phase space defined in~\tref{cuts}.~To obtain the values here we have set $A_1^{ZZ} = 2$ and all other couplings to one.~We have normalized to the (tree level) SM value for the $h\rightarrow 4\ell$ decay width corresponding to $A_1^{ZZ} = 2$ and all others zero.~{\bf Bottom:}~Same as top, but for the $4e/4\mu$ final state.}
\label{fig:absmat}
\end{figure}

As expected from our previous discussion of the $M_1 - M_2$ spectra, for the $4e$ final state similar enhancements are seen in the $Z\gamma$ couplings interfering with $A_1^{ZZ}$ while the size of the contributions from the $\gamma\gamma$ couplings remain largely unchanged as compared to when using CMS-tight cuts.~We see also that even with Relaxed-$\Upsilon$ cuts, the interference between the $\gamma\gamma$ couplings and $A_1^{ZZ}$ is still larger than for $2e2\mu$ and especially in the case of $A_2^{\gamma\gamma}$.~This implies we still have stronger sensitivity to these couplings in the $4e/4\mu$ than in $2e2\mu$, though the difference is much less drastic than when CMS-tight cuts are used.~Note however these integrated magnitudes only give a rough picture of the expected sensitivity which is achievable utilizing the fully differential cross section.

Of course the discussion so far has assumed a background free pure signal sample.~If the LHC detectors had perfect energy resolution the signal region would essentially be a delta function centered at the Higgs mass leading to an effectively background free sample.~However, detector resolution has the effect of widening the signal region, thus introducing more background into the sample.~Still, the current LHC $4\ell$ analyses provide a signal rich event sample and neglecting backgrounds is a reasonable rough approximation.~However, we have now relaxed the analysis cuts, bringing in more non-Higgs backgrounds.~It is thus important to consider the effects these backgrounds have on the sensitivity and this is the goal of the next section.

\section{Effects of Non-Higgs Background}

As mentioned above, the imperfect detector resolution has the effect of introducing non-Higgs background events into the signal region.~In essence, the resolution effects `smear' the four lepton invariant mass spectrum altering the ideal spectrum of a delta function for the signal into a gaussian-like spectrum with a $\sim 1-3$~GeV width~\cite{Khachatryan:2014kca}, where the smearing is less for muons than electrons.~A proper treatment of this spectrum requires that we combine the production mechanism with the decay for both the signal and background.~We now breifly describe how this is incorporated into our analysis, but many more details can be found in~\cite{Gainer:2011xz,Chen:2013ejz,Chen:2014pia}.

\vspace*{-.2cm}
\subsection{Signal Plus Background Likelihood}

The dominant (non-Higgs) background comes from the continuum $q\bar{q}\to4\ell$ process~\cite{Khachatryan:2014kca}.~In our analysis we include the leading order parton level fully differential cross section for $q\bar{q}\to4\ell$ which was computed analytically in~\cite{Chen:2012jy,Chen:2013ejz}.~These analytic expressions contain all possible interference effects and both the t-channel and s-channel contributions.~Following the procedure in~\cite{Gainer:2011xz}, this parton level differential cross section is then combined with the (CTEQ6l1~\cite{Lai:1999wy,Pumplin:2002vw}) initial state quark parton distribution functions (\emph{pdfs}) and `symmetrized' to account for the inability to know the incoming quark direction at a $pp$ collider such as found at the LHC.~The entire procedure is validated~\cite{Chen:2013ejz,Chen:2014pia} against Madgraph~\cite{Alwall:2011uj} over a large phase space in the range $75-1000$~GeV for the four lepton invariant mass.

The result for the four lepton invariant mass spectrum is shown in~\fref{mhspec} where our (mostly) analytic result is shown in black and the spectrum generated by Madgraph is shown in red.~We have also separated the $q\bar{q}\to4\ell$ background into its various components to see how the composition changes as a function of energy.~We see that around $\sim 125$~GeV the background is dominated by the t-channel $q\bar{q} \to Z\gamma \to 4\ell$ component (gold) followed by the s-channel $q\bar{q}\to Z\to4\ell$ (green) component both of which are much larger than the t-channel $q\bar{q} \to \gamma\gamma \to 4\ell$ (red) and $q\bar{q} \to ZZ \to 4\ell$ (blue) components.~This leads us to suspect that including the non-Higgs background will have the largest effect on the sensitivity to the $hZ\gamma$ couplings and indeed this will turn out to be the case.
\begin{figure}[tbh]
\includegraphics[width=.45\textwidth]{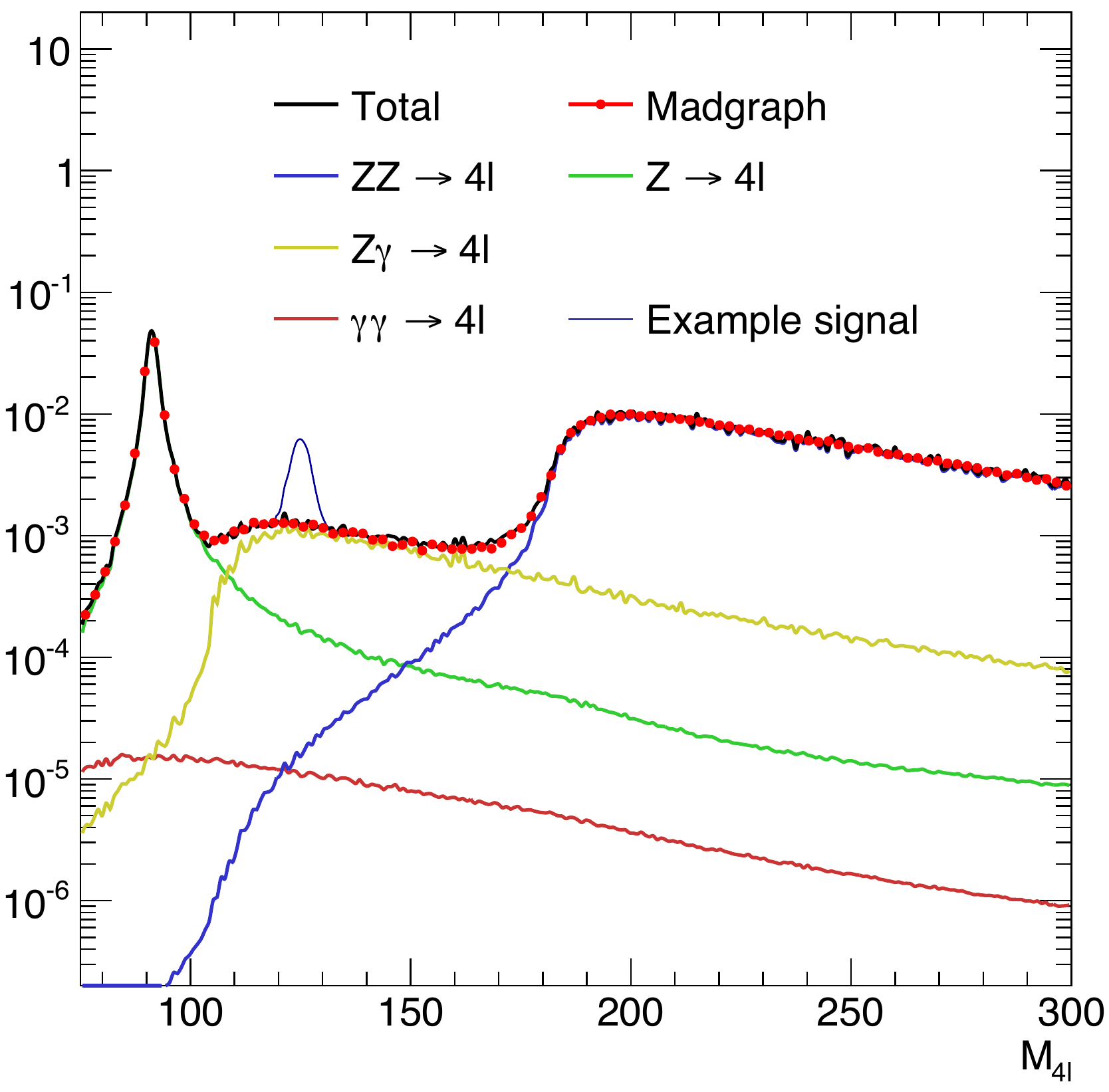}
\caption{The four lepton invariant mass spectrum for the $q\bar{q}\to4\ell$ background including \emph{pdfs}.~We plot the total background (black) and compare it to the result from a large Madgraph sample (red dots) over the range $75-300$~GeV.~We also plot the individual components which include:~t-channel $q\bar{q}\to ZZ\to4\ell$ (blue), $q\bar{q}\to Z\gamma\to4\ell$ (gold), $q\bar{q}\to \gamma\gamma\to4\ell$ (red) and s-channel $q\bar{q}\to Z\to4\ell$ (green).~The $gg\to h\to 4\ell$ signal is also shown where the Higgs peak is given a $\sigma$ of $2$~GeV and centered at $125$~GeV.}
\label{fig:mhspec}
\end{figure}

Similarly for the signal we combine the analytic expression for the $h\to4\ell$ decay~\cite{Chen:2012jy,Chen:2013ejz} with \emph{pdfs} for the $gg\to h$ production mode following the procedure in~\cite{Gainer:2011xz}.~To model the detector resolution we have smeared the signal $M_{4\ell}$ distribution with a gaussian of $\sigma = 2$~GeV centered at the Higgs mass which we take to be $125$~GeV.~Note that these resolution effects also enter into the $M_1$ and $M_2$ invariant masses.~We also plot this gaussian signal on top of the $q\bar{q}\to4\ell$ background in~\fref{mhspec}.~The complete signal plus background likelihood is then constructed as detailed in~\cite{Chen:2013ejz,Chen:2014pia} for the four lepton invariant mass window of $115-135$~GeV.~Note that the likelihoods for all $4\ell$ final states must be constructed and combined into one likelihood.~Furthermore, along with floating the six parameters in~\eref{Aall}, we must now also float the background fractions simultaneously thus accounting for correlations between the couplings and background fractions as discussed in~\cite{Chen:2013ejz,Chen:2014pia}.~We also mention that in this analysis we are utilizing a simplified implementation of detector resolution effects instead of the full detector level treatment as done in~\cite{Chen:2014pia,Khachatryan:2014kca,Chen:2014hqs}.~Since we are not precisely quantifying the sensitivity or performing a true parameter extraction, we find this simplified approach to be sufficient for present purposes.

\vspace*{-.2cm}
\subsection{Background Effects on Sensitivity}

With the signal plus background likelihood in hand we can go on to assess the effects of the $q\bar{q}\to4\ell$ background.~We see this in~\fref{SpBcurves} where we show sensitivity curves which compare the results obtained assuming a pure signal sample (solid) versus a signal plus background (dashed) sample fitting to a true point of $\vec{A} = (0,0,0,0,0,0)$.~We do this for both the CMS-tight cuts (blue) and the Relaxed$-\Upsilon$ cuts (red).~In the left plot we show the results for $A_2^{Z\gamma}$ and on the right we show $A_2^{\gamma\gamma}$.~We can see clearly that as expected the inclusion of the $q\bar{q}\to4\ell$ background has a much larger effect on the sensitivity to the $hZ\gamma$ couplings than $h\gamma\gamma$.
\begin{figure*}[tbh]
\begin{center}
\includegraphics[width=.45\textwidth]{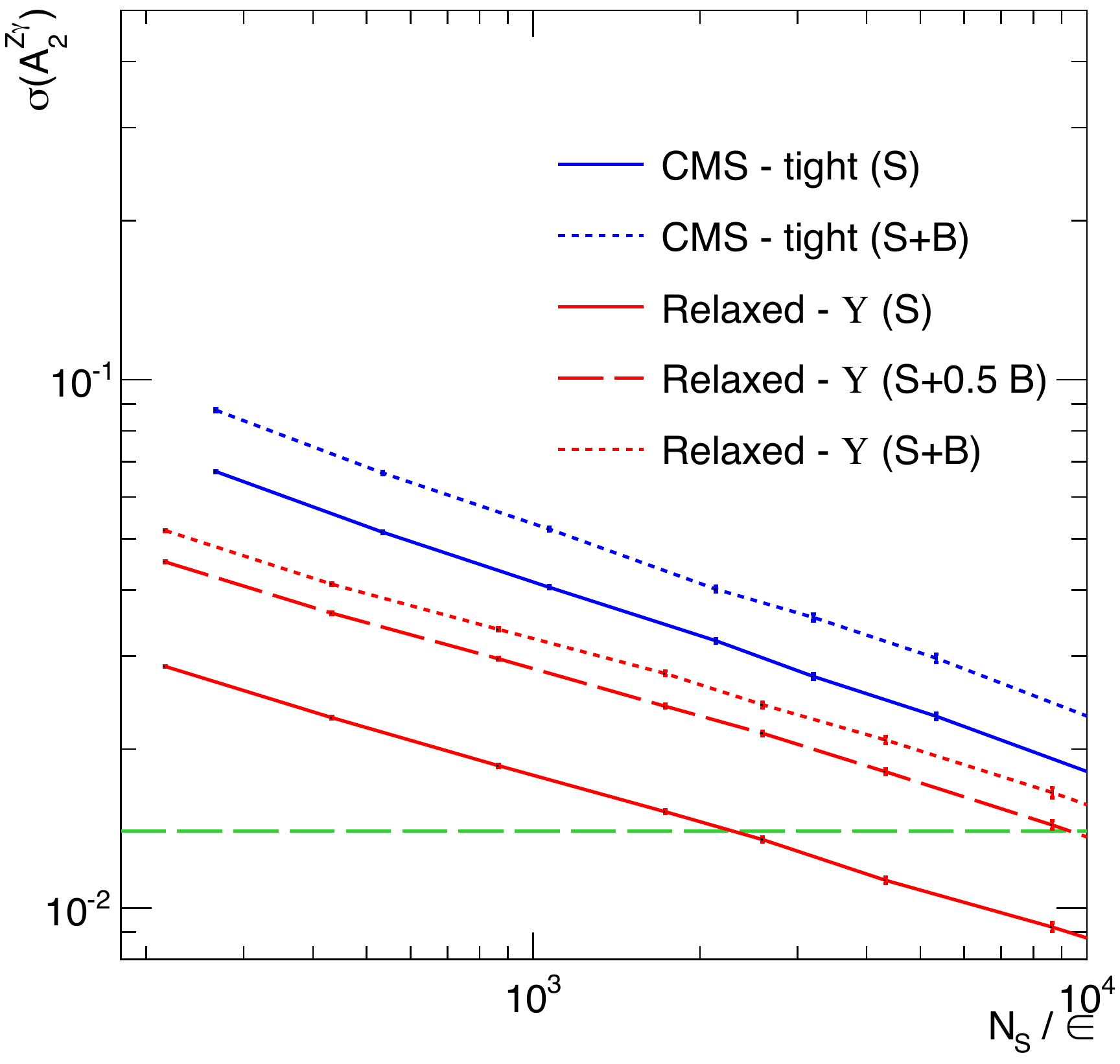}
\includegraphics[width=.452\textwidth]{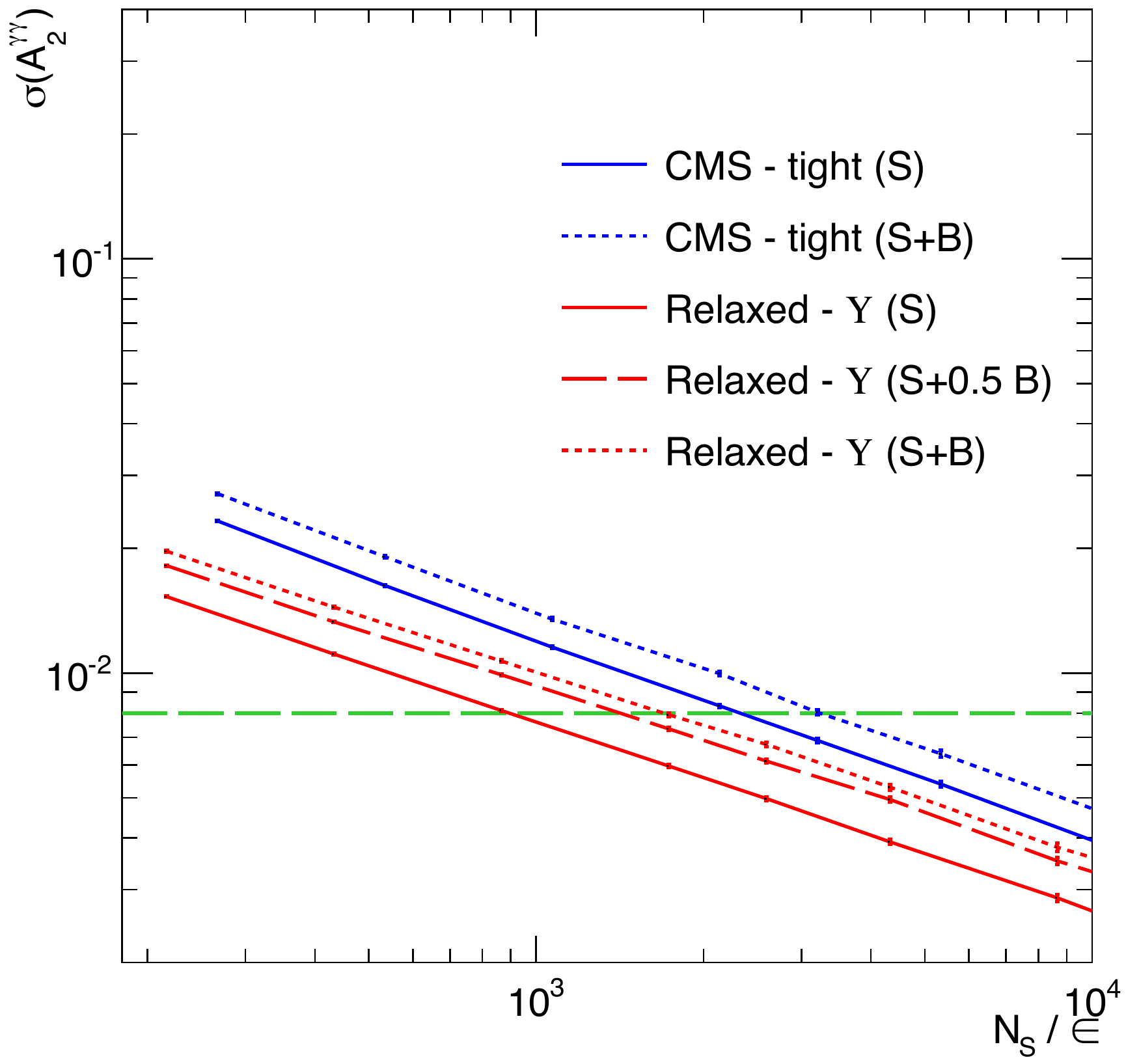}
\end{center}
\caption{Sensitivity curves for $A_2^{Z\gamma}$ (left) and $A_2^{\gamma\gamma}$ (right) as a function of number of signal events ($N_S$) divided by efficiency ($\epsilon$) for the combined $2e2\mu, 4e, 4\mu$ channels comparing pure signal (solid) versus signal plus background (dashed).~We do this for both CMS-tight cuts (blue) and Relaxed$-\Upsilon$ cuts (red) and again we have fit to a true point of $\vec{A} = (0,0,0,0,0,0)$, but now also true background fractions as indicated in~\tref{cuts}~\cite{CMS-PAS-HIG-14-014,Khachatryan:2014kca}.~We float all couplings in~\eref{Aall} as well as background fractions simultaneously to capture any potential correlations.~The green dashed lines indicate the magnitudes for the SM value of $|A_2^{Z\gamma}| = 0.014$ and $A_2^{\gamma\gamma} = 0.008$~\cite{Low:2012rj}.}
\label{fig:SpBcurves}
\end{figure*}

\emph{\underline{Background effects on $hZ\gamma$}}:~More specifically, we see that for the Relaxed$-\Upsilon$ cuts, the sensitivity to $hZ\gamma$ is degraded to the point where now $\gtrsim 10000$ events are needed to begin probing these couplings as opposed to only $\gtrsim 2000$ being needed in the pure signal case.~Interestingly, the sensitivity using the CMS-tight cuts is not as greatly affected by the presence of background.~This is because the CMS cuts are optimized to give a large signal to background ratio (see~\tref{cuts}) and thus the efficiency for background events is significantly lower than in the case of Relaxed-$\Upsilon$ cuts.~Even still, by utilizing the Relaxed$-\Upsilon$ cuts, probing these couplings may be possible towards the end of a high luminosity LHC, which is a drastic improvement over the standard CMS cuts for which $> 30,000$ events would be needed when including background. 

\emph{\underline{Background effects on $h\gamma\gamma$}}:~For the Higgs couplings to photons we see that the background again degrades the sensitivity when utilizing the Relaxed$-\Upsilon$ cuts, though not as drastically as for $Z\gamma$.~In particular, when utilizing Relaxed$-\Upsilon$ cuts, we see that in the presence of background we now need $\sim 1500-1800$ events to probe SM values, whereas in the case of pure signal only $\sim 900$ events were needed.~Again we see that for CMS-cuts the effects of background are less drastic, but still $> 3000$ events are needed which again demonstrates the improvement in sensitivity gained by using the Relaxed$-\Upsilon$ cuts.

These results demonstrate the degrading effects that the $q\bar{q}\to4\ell$ background has on the sensitivity to these couplings.~As mentioned, these enter essentially because of detector resolution effects.~As a further investigation of this, we have also performed a fit with half of the amount of background, still including a gaussian of $\sigma = 2$~GeV and find that $\sim 9000$ are now needed with Relaxed$-\Upsilon$ cuts to achieve sensitivity to $\sim$ SM values of the $hZ\gamma$ couplings.~For the $h\gamma\gamma$ the threshold is reached with $\lesssim 1400$ events.~Note that this is similar, though not equivalent to increasing the energy resolution, but gives a rough idea of the benefits of reducing the amount of background in the signal region.

The large difference between the sensitivity in the case of pure signal, which is akin to perfect detector resolution, implies that improvements in energy resolution can lead to potentially large enhancements in the sensitivity.~A more precise study of this however, requires a more in depth analysis and careful treatment of the various detector effects which are beyond the scope of our current focus though a framework for exploring these issues has been constructed in~\cite{Chen:2014pia,Chen:2014hqs}. 


\vspace*{-1.15cm}
\section{Sensitivity at the LHC}

We now move on to give an estimate of the sensitivity to all of the couplings in~\eref{Aall} at the LHC.~For this estimate we focus on the Relaxed$-\Upsilon$ cuts and include the $q\bar{q}\to4\ell$ background as well as a gaussian for the Higgs peak with $\sigma = 2$~GeV to (roughly) model the detector resolution effects.~We will first consider how the sensitivity will evolve as a function of luminosity before examining the potential to probe CP properties by the end of the LHC running with $\sim 3000 fb^{-1}$.~Results for CMS-tight cuts for pure signal can be found in~\cite{Chen:2014gka}.~To assess the sensitivity, as in all previous results shown here, pseudoexperiments are conducted on large data sets generated from a Madgraph~\cite{Alwall:2011uj} implementation of the effective operators in~\eref{dim5lag}~\cite{Chen:2013ejz,Chen:2014pia}.~Again the details of our fitting framework and procedure can be found in~\cite{Chen:2013ejz,Chen:2014pia,Chen:2014hqs,Chen:2014gka}.

\subsection{Sensitivity as Function of Luminosity}
In~\fref{LuminosityPlots} we show results for $\sigma(A_n^i)$~vs.~$N_S$ for the six couplings in~\eref{Aall} where all couplings (defined in~\eref{dim5lag}) and background fractions are floated simultaneously.~Again we have fit to a true point of $\vec{A} = (0,0,0,0,0,0)$ and background fraction as indicated in~\tref{cuts} for the range $115-135$~GeV~\cite{CMS-PAS-HIG-14-014,Khachatryan:2014kca}.~On the top axis we also indicate the luminosity $\times$ efficiency assuming a SM production (both $gg\to h$ and VBF) cross section and $h\to4\ell$ branching fraction values obtained from the LHC Higgs Cross Section Working Group~\cite{Dittmaier:2011ti,Heinemeyer:2013tqa} for a $125$~GeV mass Higgs.~We indicate by the green dashed line the value $0.008$ and the violet dashed line the value $0.014$ corresponding roughly to the magnitudes of $A_2^{\gamma\gamma}$ and $A_2^{Z\gamma}$ respectively predicted by the SM at $125~$GeV~\cite{Low:2012rj}.

We see clearly in~\fref{LuminosityPlots} the much stronger sensitivity to the $\gamma\gamma$ couplings as compared to the $Z\gamma$ and even more so the higher dimensional $ZZ$ couplings.~In particular we see that, even in the presence of the $q\bar{q}\to4\ell$ background, values of order the SM for the $h\gamma\gamma$ couplings will be probed with $\sim 100-150 fb^{-1}$ assuming $100\%$ efficiency if the Relaxed$-\Upsilon$ cuts are utilized.~Of course in a real detector $100\%$ efficiency is not achievable so a more conservative estimate is $\sim 200-500 fb^{-1}$, depending on the exact efficiency.~This allows for the exciting possibility that these couplings may be within reach of a Run-II LHC even before a high luminosity upgrade.~We see also that the sensitivity is equally strong for the CP even and odd couplings in the case of $h\gamma\gamma$ indicating sensitivity to the CP properties and potential CP violation.
\begin{figure}[h]
\begin{center}
\includegraphics[width=.45\textwidth]{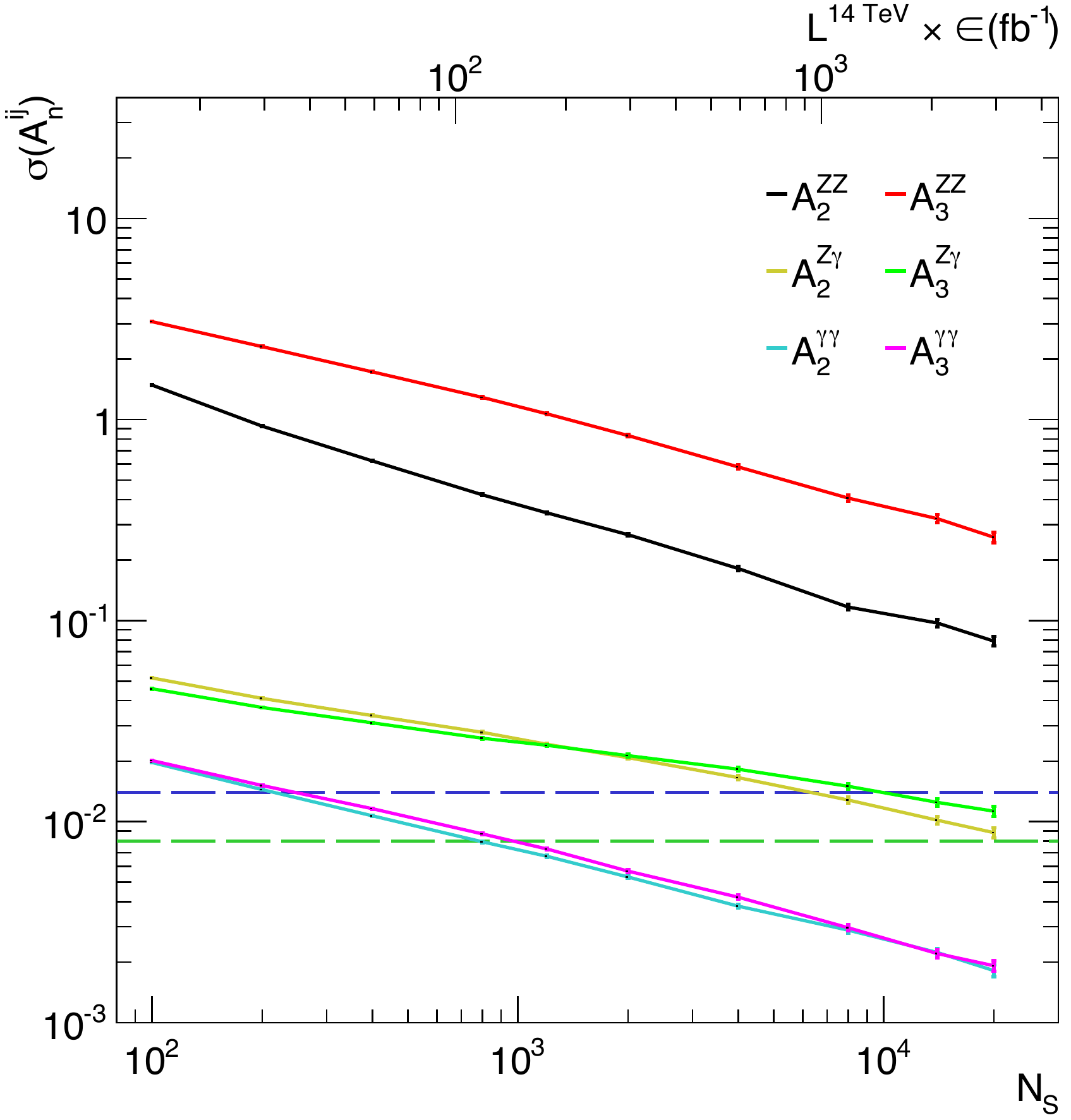}
\end{center}
\caption{$\sigma(A_n^i)$~vs.~$N_S$ for each coupling in~\eref{Aall} utilizing Relaxed$-\Upsilon$ cuts including the $q\bar{q}\to4\ell$ background for the combined $2e2\mu$, $4e$, and $4\mu$ channels.~On the top axis we also show an approximate projection for the luminosity $\times$ efficiency needed at the LHC to obtain a given number of signal events assuming SM production cross section and branching fraction values obtained from the LHC Higgs Cross Section Working Group~\cite{Dittmaier:2011ti,Heinemeyer:2013tqa}.~We indicate by the green dashed line the value $0.008$ and the pink dashed line the value $0.014$ corresponding roughly to the magnitude of $A_2^{\gamma\gamma}$ and $A_2^{Z\gamma}$ respectively as predicted by the SM at $125~$GeV~\cite{Low:2012rj}.~All couplings (defined in~\eref{dim5lag}) and background fractions are floated simultaneously and we have fit to a true point of $\vec{A} = (0,0,0,0,0,0)$ and background fraction as indicated in~\tref{cuts} for the range $115-135$~GeV~\cite{CMS-PAS-HIG-14-014,Khachatryan:2014kca}.}
\label{fig:LuminosityPlots}
\end{figure}

For the $hZ\gamma$ couplings the situation is less optimistic, but perhaps still promising at the LHC.~In particular we see that $\sim$ SM values will begin to be probed with $\sim 1000 fb^{-1}$ again assuming $100\%$ efficiency.~More realistically $2000-5000 fb^{-1}$ will likely be needed once efficiencies are accounted for.~This may still perhaps be within reach of a high luminosity LHC and certainly should be within reach of a future higher energy hadron collider.~Again we see a similar, though not identical, sensitivity to the CP even and CP odd couplings allowing for the possibility to directly probe the CP properties and potential CP violation in the $Z\gamma$ couplings.

We also see in~\fref{LuminosityPlots} that the sensitivity to the higher dimensional $ZZ$ couplings is relatively weak requiring $\sim 3000 fb^{-1}$, assuming $100\%$ efficiency, to probe couplings of $\mathcal{O}(0.08-0.09)$ for the CP even coupling and $\mathcal{O}(0.2-0.3)$ for the CP odd coupling.~This is significantly larger than what would be expected from loop effects which might generate these couplings in the SM or in most BSM extensions.~The large difference in sensitivity between the CP odd and even couplings can be understood from the fact that the sensitivity is driven by interference effects with the tree level SM $hZ^\mu Z_\mu$ operator~\cite{Chen:2014gka}.~For the $ZZ$ couplings this interference is an order of magnitude larger for the CP even operator ($A_2^{ZZ}$) than for the CP odd operator ($A_3^{ZZ}$) in contrast to the case of $\gamma\gamma$ and $Z\gamma$ where the size of the interference is of the same order for the CP odd and even couplings as can be seen in~\fref{absmat}.

\vspace*{-.25cm}
\subsection{Probing CP Properties in $hZ\gamma$ and $h\gamma\gamma$}

The results in~\fref{LuminosityPlots} indicate that the LHC may be able to directly probe the CP properties of the $hZ\gamma$ and especially $h\gamma\gamma$ couplings even for values close to the SM prediction.~This is especially exciting since there is presently no other direct probe of the CP properties of these couplings (with the possible exception of $h\to2\ell\gamma$ decays~\cite{Chen:2014ona}).~To further investigate this we perform a second fit, but now to a true point $\vec{A} = (0,0,0.014,0,-0.008,0)$ corresponding to the SM values for $A_{2,3}^{Z\gamma}$ and $A_{2,3}^{\gamma\gamma}$ at 1-loop and $125$~GeV~\cite{Low:2012rj}.~We again include the $q\bar{q}\to4\ell$ background while floating all couplings and background fractions simultaneously.~Instead of the sensitivity curves however, we examine in~\fref{ZAmoneyplot} the $1\sigma$ confidence interval for $A_2^{V\gamma}$ vs.~$A_3^{V\gamma}$ ($V = Z, \gamma$) couplings utilizing CMS-tight (large, light turquoise ellipse) and Relaxed$-\Upsilon$ (small, dark turquoise ellipse) cuts assuming $4000$ events corresponding roughly to $3000 fb^{-1}$~\cite{Dittmaier:2011ti,Heinemeyer:2013tqa}.~For comparison and as a demonstration of the ideal case, we also show in the red ellipses the $1\sigma$ interval obtained assuming a pure signal sample.

\emph{\underline{Probing CP properties of $hZ\gamma$}}:~On the left in~\fref{ZAmoneyplot} we show results for the sensitivity of our analysis in the $A_2^{Z\gamma}$ vs.~$A_3^{Z\gamma}$ plane.~We also indicate by the pink rings the projected $1\sigma$ interval from the on-shell $h\to Z\gamma$ decay rate for $3000 fb^{-1}$ respectively~\cite{htoAA}.~Our true point is represented by the star at $(A_2^{Z\gamma}, A_3^{Z\gamma}) = (0.014, 0)$.~In~\fref{ZAmoneyplot} one can see clearly the improvement in sensitivity one obtains using the Relaxed-$\Upsilon$ cuts versus standard CMS-tight cuts.~Qualitatively we see that in the case of Relaxed-$\Upsilon$ cuts almost the entire $1\sigma$ region lies on the positive side of zero for $A_2^{Z\gamma}$ indicating that with these cuts the LHC has a better chance to establish the overall sign of the $A_2^{Z\gamma}$ coupling than with the standard CMS cuts and something which can not be done in $h\to Z\gamma$ on-shell two body decays.~One can quantify this further by taking the ratio of the area corresponding to the CMS-tight $1\sigma$ ellipse over the corresponding one for Relaxed-$\Upsilon$ cuts.~For the ellipses in~\fref{ZAmoneyplot} corresponding to $\sim 3000fb^{-1}$ we find this ratio to be $\sim 2.2$ implying a $\sim 120\%$ improvement.~We also notice the asymmetric nature of the ellipses indicating a somewhat stronger sensitivity to the CP even coupling than for the CP odd as already implied by the sensitivity curves in~\fref{LuminosityPlots}.~As a reference, the ideal case of pure signal is also shown in red and gives a clear indication of the degrading effects due to detector resolution which introduces non-Higgs background into the signal region.
\begin{figure*}
\begin{center}
\includegraphics[width=.435\textwidth]{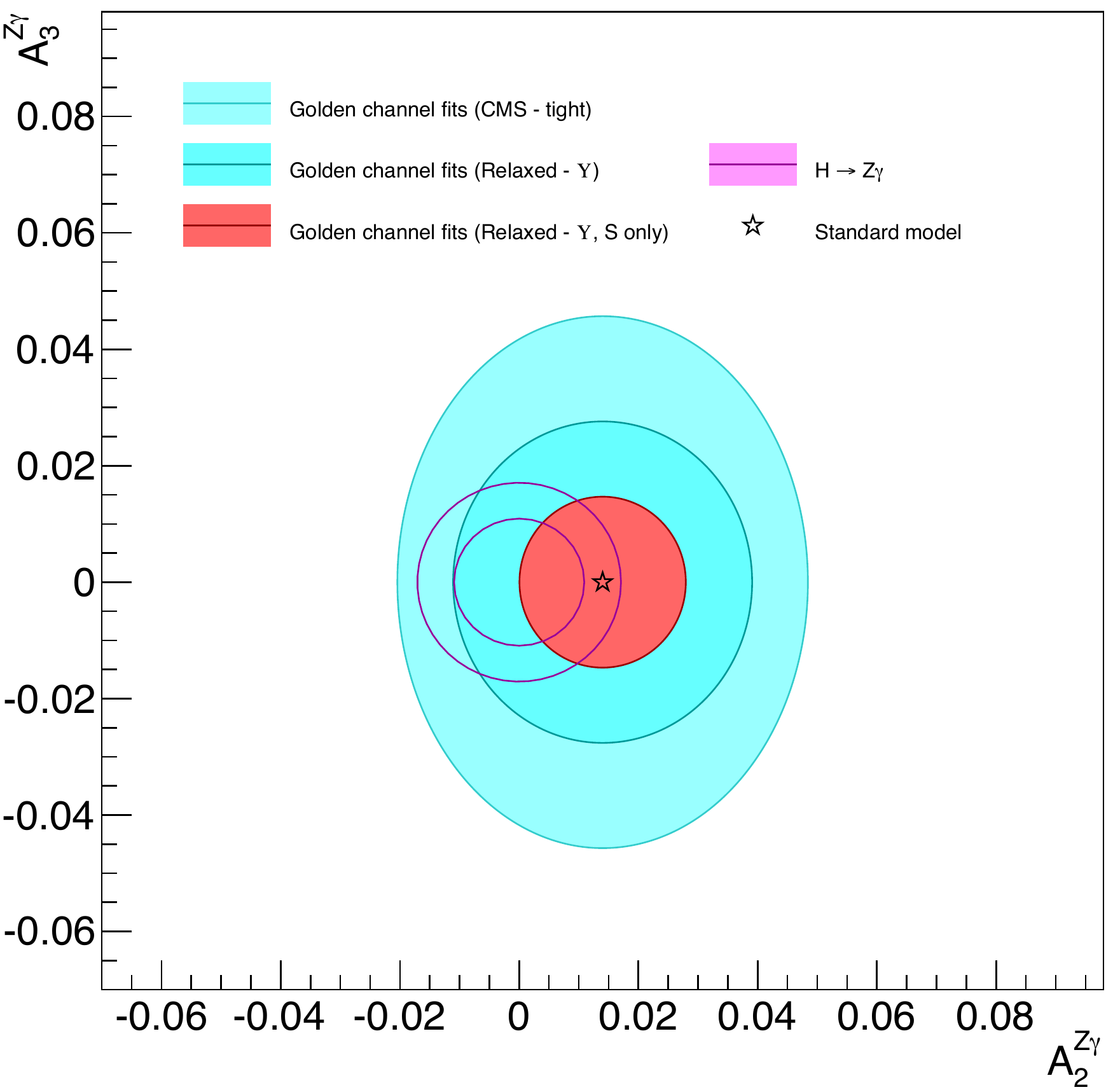}
\includegraphics[width=.45\textwidth]{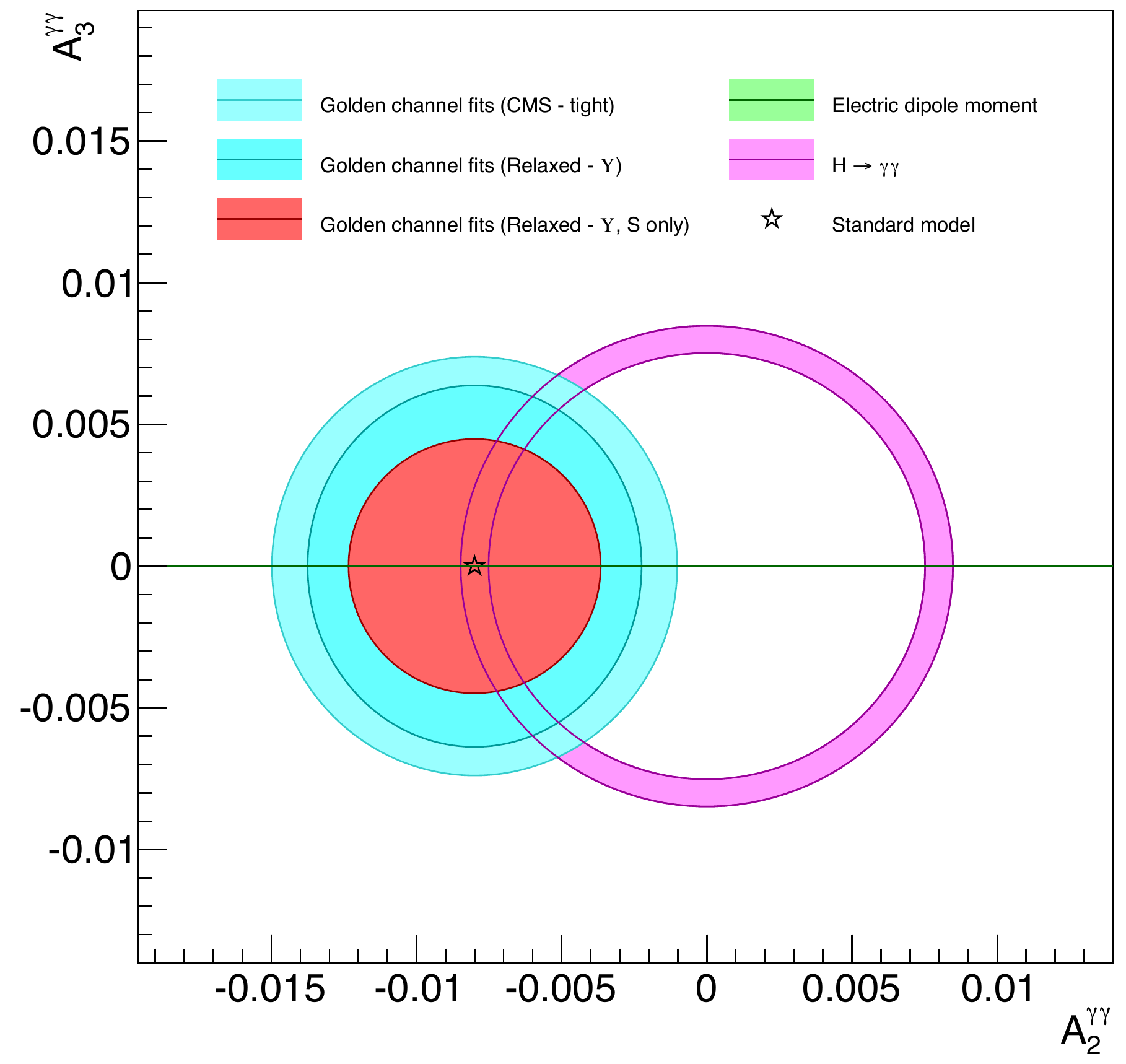}
\end{center}
\caption{{\bf Left:}~Results for the for $A_2^{Z\gamma}$~vs.~$A_3^{Z\gamma}$ assuming $4000$ events corresponding to roughly $3000 fb^{-1}$~\cite{Dittmaier:2011ti,Heinemeyer:2013tqa} (after accounting for efficiencies).~The same fit as in~\fref{LuminosityPlots} is performed only we fit to a true point of $\vec{A} = (0,0,0.014,0,-0.008,0)$ represented by the star and corresponding to the SM values for $A_2^{Z\gamma}$ and $A_2^{\gamma\gamma}$ at $125$~GeV~\cite{Low:2012rj}.~The turquoise ellipses correspond to the $1\sigma$ confidence interval obtained in the golden channel for CMS-tight (large, light turquoise) and Relaxed$-\Upsilon$ (small, dark turquoise).~The pink ring indicates the projected $1\sigma$ confidence interval which will be achieved on the $h\to Z\gamma$~\cite{htoAA} rate for the same luminosity.~We also show in the red ellipse the projected sensitivity assuming a pure signal sample.~{\bf Right:}~Same as in left, but for $A_2^{\gamma\gamma}$~vs.~$A_3^{\gamma\gamma}$.~We also include a thin green line showing the severe, but model dependent constraint coming from the electron EDM in a minimal model where the mass of the states which generate these operators is a TeV and that the Higgs couplings to first generation fermions are of order their SM value~\cite{McKeen:2012av,Baron:2013eja}.}
\label{fig:ZAmoneyplot}
\end{figure*}

\emph{\underline{Probing CP properties of $h\gamma\gamma$}}:~On the right in~\fref{ZAmoneyplot} we show results for the $A_2^{\gamma\gamma}$ vs.~$A_3^{\gamma\gamma}$ couplings, again comparing to the projected sensitivity for the on-shell decay.~Here we also include a thin green line showing the severe, but model dependent constraint coming from the electron EDM in a minimal model where the mass of the states which generate these operators is a TeV and that the Higgs couplings to first generation fermions are of order their SM value~\cite{McKeen:2012av,Baron:2013eja}.~The true point is again represented by the star, but now at $(A_2^{\gamma\gamma}, A_3^{\gamma\gamma}) = (-0.008, 0)$.~We see clearly that the overall sensitivity is much stronger for the $\gamma\gamma$ couplings than for $Z\gamma$ making it clear that the overall sign of the $A_2^{\gamma\gamma}$ should be established at the LHC regardless of cuts used.~However again we see a significant improvement in sensitivity is found when utilizing Relaxed$-\Upsilon$ versus CMS-tight cuts although it is not as drastic as for the $Z\gamma$ couplings.~Taking the ratio of the areas again we find $\sim 1.4$ indicating $\sim 40\%$ improvement.~We also note the symmetric nature of the ellipses now further exemplifying the equal sensitivity to both the CP odd and even couplings.~The ideal case of pure signal is shown in red where we see once again that background effects degrade the sensitivity though not as drastically as for the $hZ\gamma$ couplings.

We also note once again that the sensitivities obtained here may be enhanced further by including the regions around the $\Upsilon$ mass and below $M_{1,2} \sim 4$~GeV which would require proper treatment of the various QCD resonances as well as large $Z-\gamma$ mixing effects~\cite{Gonzalez-Alonso:2014rla}.~Due to the strong discriminating power in these regions, their inclusion may bring the luminosities needed to probe the $\gamma\gamma$ couplings to well within reach of Run-II and the $Z\gamma$ couplings to well within reach of a high luminosity LHC.~However we leave an investigation of this to future work. 

\subsection{Beyond the LHC}

A future hadron collider will have the advantage over the LHC of much larger $h\to4\ell$ event rates due to the large production cross sections and in particular for $gg\to h$.~To get an idea of what can be achieved with these larger data sets we show in~\fref{VAmoneyplots} the same plots as in~\fref{ZAmoneyplot}, but for $20$k events which should be well within reach of a future hadron collider operating at $33$ or $100$~TeV.~For the $1\sigma$ projections on the $h\to Z\gamma$ and $h\to\gamma\gamma$ rates we assume the progression is purely statistical and rescale the projections for 3000~$fb^{-1}$ accordingly.~These results imply a future machine should drastically improve the sensitivity and the potential to discover new physics such as CP violation in these couplings. 
\begin{figure*}
\begin{center}
\includegraphics[width=.445\textwidth]{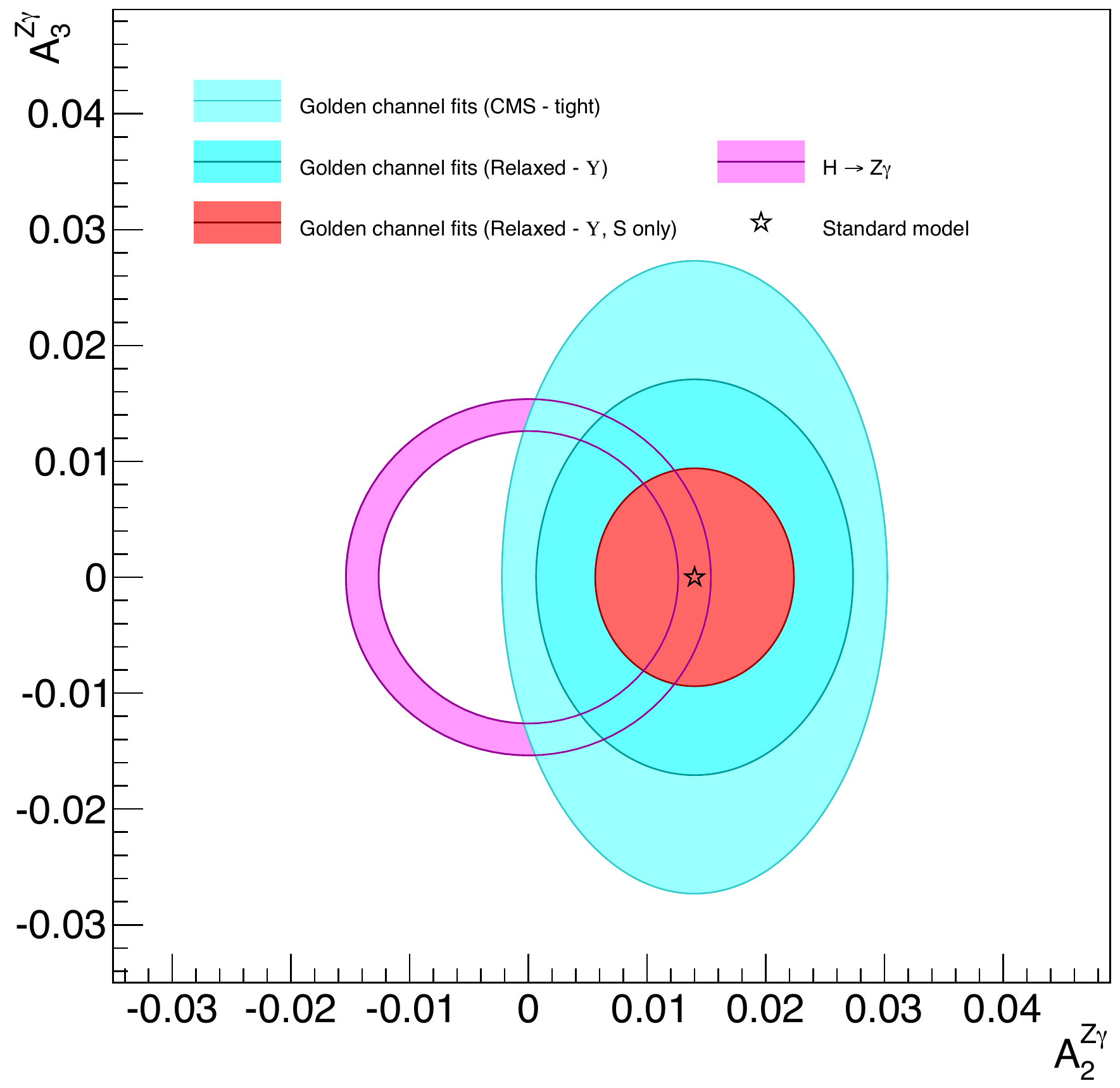}
\includegraphics[width=.45\textwidth]{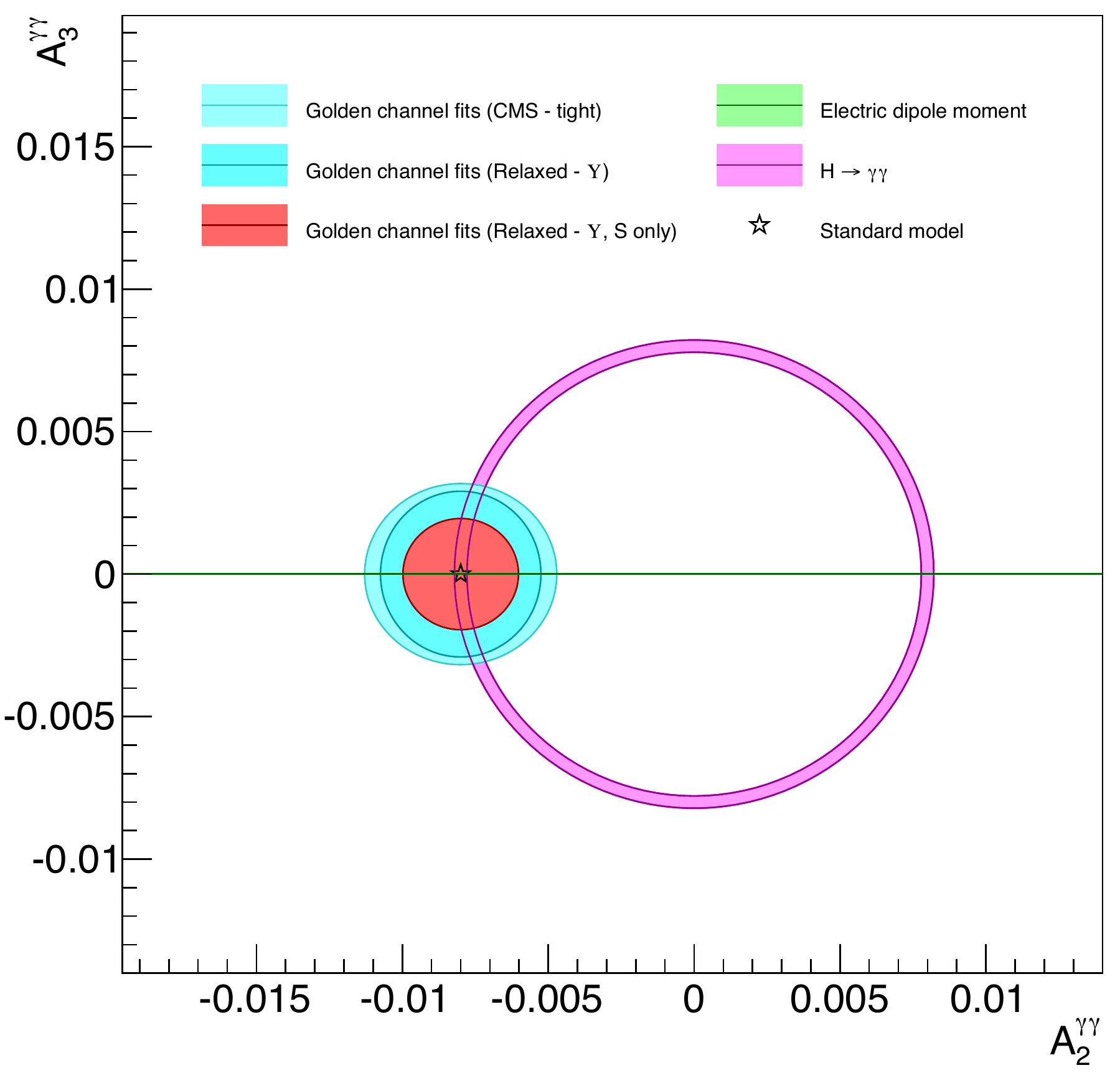}
\end{center}
\caption{Same as~\fref{ZAmoneyplot}, but for 20,000 events which should be well within reach of a future high energy hadron collider.~On the left we show $A_2^{Z\gamma}$~vs.~$A_3^{Z\gamma}$ and on the right $A_2^{\gamma\gamma}$~vs.~$A_3^{\gamma\gamma}$.~The pink rings indicate the projected $1\sigma$ confidence interval which will be achieved on the $h\to Z\gamma$ and $h\to \gamma\gamma$ rates obtained by simply rescaling the projections at 3000~$fb^{-1}$~\cite{htoAA}.}
\label{fig:VAmoneyplots}
\end{figure*}

\section{Conclusions}
\label{sec:conclusion}

We have performed an analysis of the expected sensitivity in the $h\to4\ell$ channel to the higher dimensional Higgs couplings to $ZZ$, $Z\gamma$, and $\gamma\gamma$ pairs.~To do this we have utilized a framework based on analytic expressions for the $h\to4\ell$ signal and dominant $q\bar{q}\to4\ell$ background fully differential cross sections in order to perform a multi-dimensional parameter extraction.

We have demonstrated that utilizing relaxed cuts or alternative lepton pairings during event selection can significantly enhance the sensitivity of the $h\to4\ell$ channel to the Higgs couplings to $Z\gamma$ and $\gamma\gamma$ pairs relative to that found utilizing current CMS event selection criteria.~In particular we have proposed a set of relaxed cuts which give a $\gtrsim 100\%$ enhancements in sensitivity to the CP properties of the $hZ\gamma$ couplings and $\gtrsim 40\%$ enhancements for the $h\gamma\gamma$ couplings.

With this enhancement we estimate that the sensitivity to the $h\gamma\gamma$ couplings begins to reach the levels necessary to probe values of order the Standard Model prediction with $\sim 200-500 fb^{-1}$ depending on detector efficiencies, perhaps within reach of a Run-II LHC and certainly a high luminosity LHC.~For the Higgs couplings to $Z\gamma$ we estimate that $\sim 2000-5000 fb^{-1}$ will be needed allowing them to perhaps be probed at a high luminosity LHC and certainly at a future high energy hadron collider.~We have also discussed the fact that the results obtained here can in principle be improved upon by relaxing the cuts even further and/or improving detector energy resolution.~We leave a more detailed study of further optimization and possibilities at a 100~TeV collider to future work.

These direct measurements of the $h\gamma\gamma$ and $hZ\gamma$ CP properties can not be made in the $h\rightarrow \gamma\gamma$ and $h\rightarrow Z\gamma$ on-shell two body decay channels or in other indirect approaches without making model dependent assumptions.~This makes the $h\to4\ell$ golden channel the unique method capable of determining these properties in the foreseeable future and we encourage experimentalists at the LHC to carry out these measurements.
\\~\\
\noindent
{\bf Acknowledgments:}~We thank Joe Lykken and Maria Spiropulu for providing us with the resources necessary to complete this study and Adam Falkowski for useful comments on the manuscript.~We also thank Ian Low, Javi Serra, and Daniel Stolarski for helpful discussions.~R.V.M.~is supported by the ERC Advanced Grant Higgs@LHC.~Fermilab is operated by Fermi Research Alliance, LLC, under Contract No.~DE-AC02-07CH11359 with the United States Department of Energy.~Y.C.~is supported by the Weston Havens Foundation and DOE grant No.~DE-FG02-92-ER-40701.~This work is also sponsored in part by the DOE grant No.~DE-FG02-91ER40684 and used the Extreme Science and Engineering Discovery Environment (XSEDE), which is supported by National Science Foundation grant number OCI-1053575.


\bibliographystyle{apsrev}
\bibliography{GoldenChannelBib}

\end{document}